\documentclass{vgtc}                          

\ifpdf
  \pdfoutput=1\relax                   
  \usepackage{graphicx}                
  \DeclareGraphicsExtensions{.pdf,.png,.jpg,.jpeg} 
\else
  \usepackage{graphicx}                
  \DeclareGraphicsExtensions{.pdf,.eps}     
\fi%

\graphicspath{{figures/}{pictures/}{images/}{./}} 

\usepackage{microtype}                 
\PassOptionsToPackage{warn}{textcomp}  
\usepackage{textcomp}                  
\usepackage{mathptmx}                  
\usepackage{times}                     
\usepackage{cite}                      
\usepackage{tabu}                      
\usepackage{booktabs}                  

\usepackage{amsmath}
\usepackage{color}
\usepackage{todonotes}
\usepackage{url}

\onlineid{0}

\vgtccategory{Research}

\vgtcinsertpkg

\title{Visualization and Comparison of AOI Transitions \\with Force-Directed Graph Layout}

\author{Yuri Miyagi\thanks{e-mail: miyayuri.25@gmail.com}\\ %
        \scriptsize Ochanomizu University %
\and Nils Rodrigues\thanks{e-mail: nils.rodrigues@visus.uni-stuttgart.de}\\ %
     \scriptsize University of Stuttgart %
\and Daniel Weiskopf\thanks{e-mail: daniel.weiskopf@visus.uni-stuttgart.de}\\ %
     \scriptsize University of Stuttgart %
\and Takayuki Itoh\thanks{e-mail: itot@is.ocha.ac.jp}\\ %
     \scriptsize Ochanomizu University %
     }

\abstract{
By analyzing the gaze trajectories of people viewing screens and advertisements, we can determine what people are interested in.  This knowledge can be effective when recommending commercial products and services, and also, when improving advertisement design.   Therefore, analysis and visualization of eye gaze have been an active research topic.  This paper proposes a new method for visualizing patterns of the gaze trajectories of multiple people by (1) visualizing patterns that move through multiple areas of interest (AOI) and (2) visualizing differences among multiple gaze trajectories.   The method first constructs a hierarchical AOI structure to a Web page or an image, and uses this structure to convert the trajectory into a sequence of symbols. We apply N-grams to the generated symbol sequences to extract transition patterns between AOIs.  Finally, the method visualizes a list of the pattern extraction results and the shapes of the characteristic elements.  We present the visualization of gaze trajectories for three examples of stimuli, and argue that analysts can efficiently discover trends in gaze transitions between text and figures, as well as differences between participants of the eye-tracking experiments.
} 

\let\olditemize\itemize
\renewcommand{\itemize}{
\olditemize
\setlength{\itemsep}{1.2pt}
\setlength{\parskip}{0pt}
\setlength{\parsep}{0pt}
}

\begin{document}


\firstsection{Introduction}
\maketitle

Visualization of eye-tracking data clarifies peoples' behavior, for instance: ``What did people watch?,'' ``How is the order of looking at a series of objects?,'' ``How long did people gaze the particular item?''.
The items that attracted people's attention indicate their interests.
This information is useful in the various fields, for example, the suggestion of products reflecting a person's interests and improvement of the design of Web pages and advertisements.
%
%
Visualization of scan-paths is one of the major methods to analyze such eye-tracking data. 
It has the advantage of showing specific movements of the recorded eye-tracking data.
The goal is to generate an easy-to-understand visual representation.

However, existing visualization methods lack proper support to address the following two issues.
The first problem is the lack of a way to visualize long transition patterns among multiple areas of interests (AOIs). 
Generating AOIs and finding such patterns leads to finding more details of people's behaviors and interests, for example: ``What combination of items are people interested in?'' and ``In which order do people observe several pieces of contents?''.
Visualizing such patterns as scan-paths is, however, not easy for long patterns due to complicated overlapping of lines.
%
The second issue is the comparison of multiple eye tracking scan-paths. 
Such a comparison is important for finding characteristic behaviors of a person or common viewing patterns. 
Direct visualization of such differences helps us to find what we should consider to distinguish eye tracking movement and behaviors.
Such information indicates how to classify several movements and how to find common or exceptional movement.
In most of the existing scan-path visualization methods, we generate an individual trajectory for each person and compare the visualization results. 
A direct drawing of differences of multiple scan-paths may be helpful to save time in the analysis process.

In this paper, we propose a new visualization technique for scan-paths to address the above two problems. 
We define AOIs and convert the scan-paths to sequences of characters. 
The generated strings indicate at which AOI the participant looked. 
The characteristic point of our technique is the use of hierarchical AOIs and the application of N-grams. 
N-grams are helpful to find patterns in strings that represent transitions between AOIs. 
Hierarchical AOIs can further help understand detailed features of patterns such as the relationship of AOIs that the participant observed successively. 
%

We propose two types of visualization that work in a combined fashion.
One is a stacked bar chart to show the results of pattern extraction by N-grams.
This method helps us focus on characteristic movements such as frequent actions or behavior that is unique for a particular participant.
The other is a force-directed graph layout to describe the extracted transitions between AOIs.
This graph is effective to understand the concrete shape of extracted patterns.
The combination of these two visualization components is effective focusing on characteristic patterns selectively.
Such selection of patterns leads to improving the readability of the visualization of long patterns and helps to address the first problem.
Meanwhile, we use the stacked bar chart to represent differences in contained patterns between two persons in order to solve the second problem.
%
As case studies, we applied our technique to three stimuli and analyzed patterns and compared differences of behaviors.
%
In summary, the contributions of our paper are the following:
\vspace{-0.5\baselineskip}   
\begin{itemize}
    \item Combination of the visualization of scan-paths and pattern extraction.
    \item Comparison of scan-paths collected from several participants, using hierarchical AOIs and visualization of differences.
    \item Automatic generation of fine AOIs.
\end{itemize}
\vspace{-0.5\baselineskip}

This paper is an extension our Japanese domestic paper \cite{Miyagi19} by presenting the extended system implementation and introducing more case studies.

 %
The remainder of this paper is organized as follows. 
In Section \ref{sec_related}, we discuss related work focusing on visualization of eye tracking scan-paths, techniques to analyze transitions between AOIs, and other visualization techniques. 
Section \ref{sec_proposed} presents our technique in detail, and Section \ref{sec_experiment} documents the results of case studies. 
Section \ref{sec_discussion} discusses the results shown in Section \ref{sec_experiment}, and finally, Section \ref{sec_conclusion} summarizes this paper.

\section{Related Work}
\label{sec_related}
This section discusses related work split into papers on visualization methods for eye-tracking data mainly using scan-paths and the visualization methods for transitions among AOIs. 

\subsection{Visualization of Eye Tracking between Scan-paths}
\label{sec_related_scanpaths}

The visualization of scan-paths is a way to show viewing behavior. 
There are various solution to avoid complicated visualization results for complex scan-path data. 
%
The following two techniques focus on the aggregation of trajectories for showing representative behaviors.
%
Krueger et al. \cite{Krueger16} proposed a technique for visualizing scan-paths on a static stimulus. Its distinctive feature is to enable the user to move lenses and display details of a particular region, such avoiding to show complicated paths. 
Peysakhovich et al. \cite{Peysakhovich18} visualized eye-tracking data over paintings. 
They generated scan-paths in rainbow colors on the grayscale pictures and the colors indicated directions or orders of passing each point. 
They applied edge bundling to simplify the scan-paths. 

The following two methods can generate hierarchies to classify scan-paths, and arrange what to visualize. 
Rodrigues el al. \cite{Rodrigues18} developed a method for summarizing and visualizing scan-paths. 
The authors prepared a lattice and drew scan-paths on it. 
Users could set the degree of the summarization by changing the resolution of the lattice. 
Kuebler et al. \cite{Kuebler16} visualized eye-tracking data over paintings by using a heat map and drawing scan-paths.
They divided a scan-path and applied clustering according to directions and positions.

Li et al. \cite{Li18} proposed mixing several paths and generating an average trajectory.
They generated AOIs based on distributions of scan-paths and displayed transitions using average scan-path and timeline visualization.
%
Gu et al. \cite{Gu17} proposed ETGraph which can display both specific scan-paths and results of a comparison of participants.
They firstly generated fine polygonal AOIs following distributions of scan-paths. 
Then, they lined up both a scan-path on a static stimulus and an abstract graph that can clearly show transitions between the AOIs.

Our technique differs from previous work in several respects: simplification of scan-paths, description of transitions among multiple AOIs, and selection of patterns to visualize. 
We use a force-directed graph layout to emphasize the transitions by reducing overlap of the scan-paths.
Also, we use the pattern extraction results to find differences between two scan-paths as well as representative movements.

\subsection{Visualization of AOI Transitions}
\label{sec_related_transitions}

Another class of  methods focuses on the visualization of transitions between AOIs. 
To begin with, there are different ways to define AOIs.
Roughly, AOIs tend to have a relationship to the contents of the stimulus or the distribution of scan-paths. 
A relatively simple AOI definition based on contents of the stimulus appears in a study by Holmqvist et al \cite{Kenneth03}. 
%
Burch et al. \cite{Burch14} developed three types of methods to define AOIs: the hot-spots-driven, the semantics-driven, and the grid-based method. 
They also proposed a technique for visualizing transitions between AOIs using node-link and matrix-based visualization. 
Blascheck et al. \cite{Blascheck16} proposed a composite visualization technique for hierarchical AOIs. 
Their technique is effective in analyzing characteristic movements while checking the classification of products. 
Burch et al. \cite{Burch18} proposed a method for defining AOIs from a distribution of eye tracking scan-paths. 
Users can observe transitions between the AOIs using a timeline. 
%
We define AOIs based on stimulus contents to compare multiple scan-paths on common AOIs.
Furthermore, we aim to find non-expected relationships between multiple AOIs that have different contents and seem non-related at first sight.

The following papers deal with the visualization of transitions between AOIs.
These techniques mainly use visualization by timeline charts or matrices.
Trefzger et al. \cite{Trefzger:18} compared several timeline visualization results that show transitions between several AOIs.
Yang et al. \cite{Yang18} proposed Alpscarf, which can clearly visualize the order of passing through AOIs.
The characteristic point is to focus on the visualization of whether participants read contents with a supposed order or included movement in the opposite direction.
Muthumanickam et al. \cite{Muthumanickam18} worked on visualizing changes in AOIs.
They generated AOIs based on distributions of scan-paths, i.e.,  the shapes of AOIs varied in time.
The following two papers compose multiple visualization techniques. 
Rudi et al. \cite{Rudi18} analyzed eye-tracking data of pilots on a cockpit.
They combined scan-path visualization, specific positions of AOIs, and timeline visualization of transitions among AOIs.
Blascheck et al. \cite{Blascheck19} utilized several eye tracking visualization methods such as AOI timelines and scan-paths.
Their technique showed how participants observed various visualization results such as stacked bar charts, bubble charts, and line charts.

However, such ways often cause difficulty to understand specific behavior or positions on the stimulus.
Our methods utilize scan-paths to show the positions where AOI transitions occurred.

\section{Proposed Technique}
\label{sec_proposed}

\begin{figure}[!h]
 \centering 
 \includegraphics[width=\columnwidth]{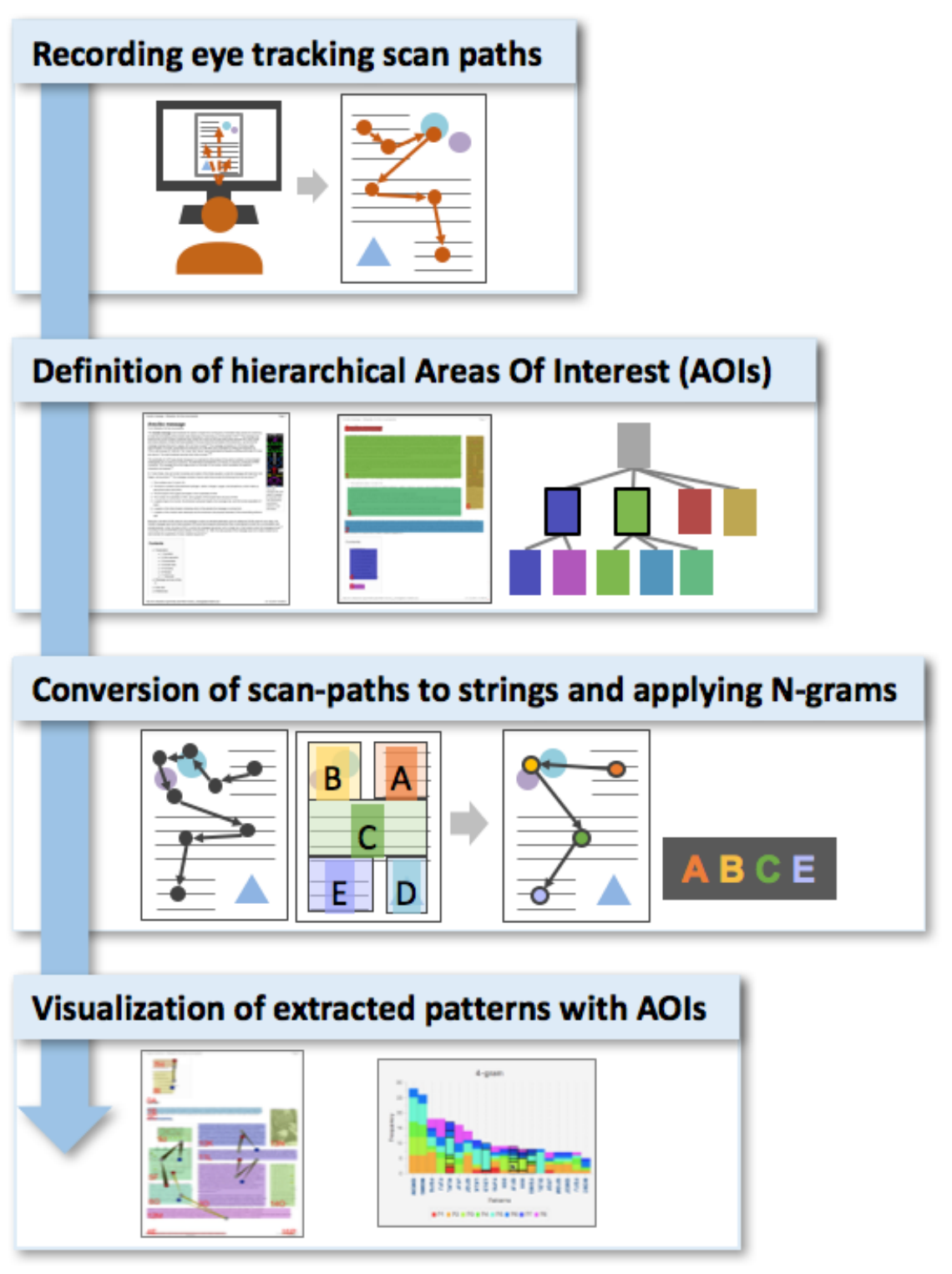}
   \caption{Processing flow of the proposed technique.}
 \label{fig_flow}
\end{figure}

We introduce the details of our technique in this section.
Figure \ref{fig_flow} illustrates the processing flow.
Section \ref{sec_preparation} describes the recording of eye tracking scan-paths and definition of hierarchical AOIs.
Section \ref{sec_generation} explains how to generate sequences of characters from the scan-paths by applying N-grams to extract patterns.
Finally, Section \ref{sec_visualization} introduces two methods to visualize the results of the pattern extraction.

\subsection{Preparation of eye-tracking data}
\label{sec_preparation}

We describe the transitions of eye tracking scan-paths between AOIs as described in Section 1.
Our implementation defines AOIs as a hierarchical structure since it is convenient to both finely and roughly describe the transitions.
This section describes the preparation steps of our technique that record the eye-tracking data and generate hierarchical AOIs.

\subsubsection{Recording eye-tracking data}
\label{sec_recording}
Our targets are scan-paths that depict the order of looking at items on a static stimulus, for instance, posters for presentation or Web pages that have multiple text fields and pictures are suitable for our technique.
We display such a stimulus on a display and ask participants to observe the picture.
We use Tobii Pro T60 XL to record scan-paths while the participant is looking at the pictures or documents.
This device can record gaze at 60 Hertz.
This frequency is high enough as we aim to find features of behavior in a relatively wide-scale such as transitions between several contents at this time.
We define the $i$-th participant as $P_{i}$ and the scan-path of $P_{i}$ that has $n_i$ points as $C_i$:

\begin{equation}
  C_{i} = \{c_{i1}\;c_{i2} \; ... \; c_{in_{i}}\}
  \label{eq:path}
\end{equation}

\subsubsection{Definition of Hierarchical AOIs}
\label{sec_definition}

\begin{figure}[!h]
 \centering 
 \includegraphics[width=\columnwidth]{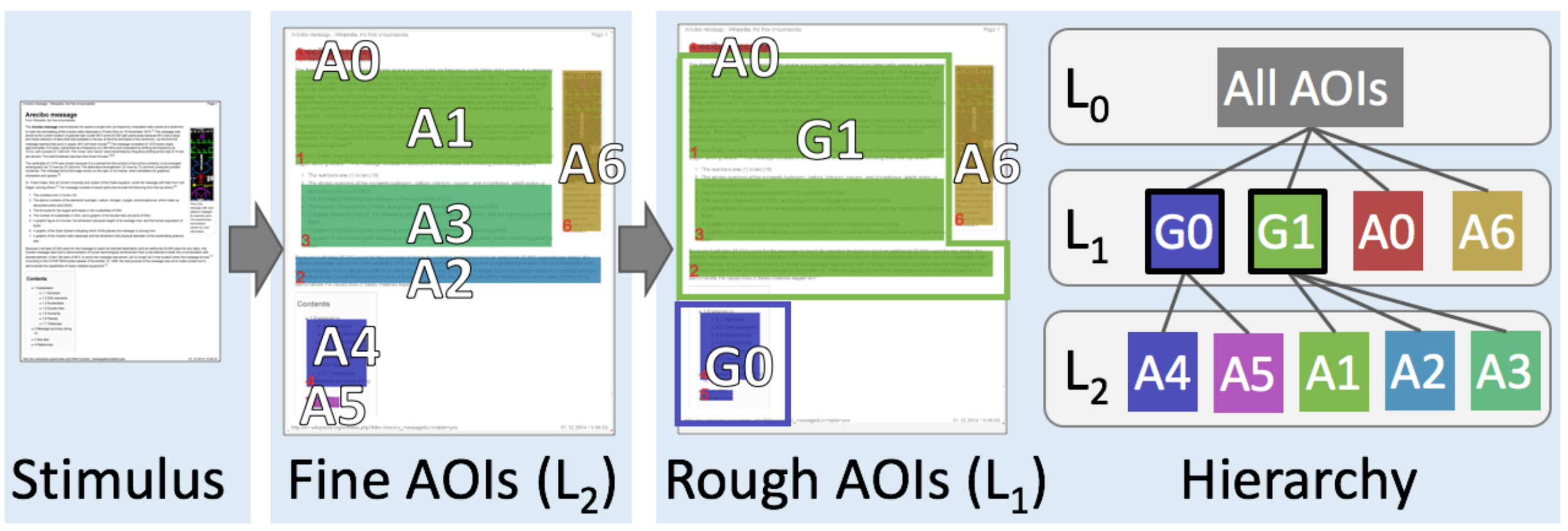}
   \caption{The flow of the definition of hierarchical AOIs.}
 \label{fig_hierarchy}
\end{figure}

The next step is the generation of hierarchical AOIs on the stimulus.
Here, an AOI is defined as a rectangle.
The process is roughly divided into two steps as shown in Figure \ref{fig_hierarchy}.
The first step is a specification of the finest AOIs reflecting content in the stimulus.
The second step is to establish coarse AOIs and a hierarchy by grouping the fine AOIs.
%
%

\begin{figure}[!h]
 \centering 
 \includegraphics[width=\columnwidth]{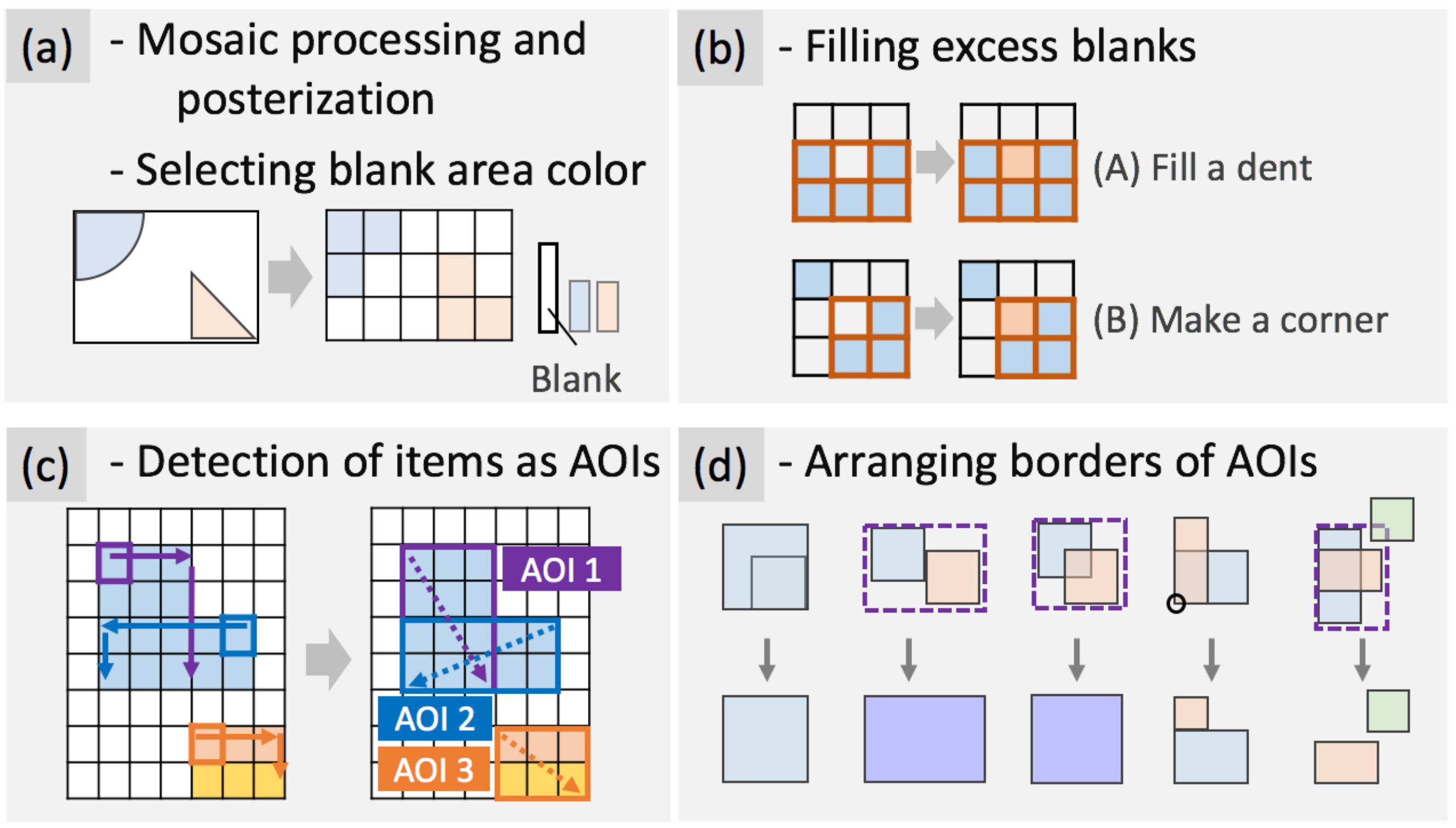}
   \caption{Process of automatic definition of AOIs.}
 \label{fig_defineaoi}
\end{figure}

In the first step, users can choose if they want manual or automatic generation of detailed AOIs.
The manual procedure is based on repeatedly dragging the pointer crossing an object corresponding to single AOI.
While repeating the AOI creation process, already created AOIs may slightly change their sizes to line up according to the newest AOI, so as to avoid overlapping of AOIs.
Users can reduce effort by choosing automatic AOI definition if they use a stimulus with a relatively simple structure.
This automatic definition uses color distribution on the stimulus to classify objects as AOIs and blank areas.
Figure \ref{fig_defineaoi} shows the flow of the automatic process.
%
%
First, we simplify contents in the static stimulus to make setting rectangular AOIs easy, as shown in Figure \ref{fig_defineaoi}(a).
The mosaic processing is helpful while grasping a rough shape of each item on the stimulus.
In the later process, borders of the AOIs will trace the borders of generated pixels.
Users can specify the sizes of the pixels $z$ in the mosaic image.
Also, we apply the color reduction process to the mosaic image.
This process improves the clarity of the borders between the items and the blank area.
Users can specify the number of colors $g$ for the color reduction.
Our implementation specifies the color of the blank areas as the color that covers the largest number of pixels, and regard other colors as the ones that depict the items corresponding to AOIs.
The next step is the transformation of the regions of item colors to rectangles, as shown in Figure \ref{fig_defineaoi}(b).
We repaint several pixels painted in the blank area color by another color to reduce excess blank areas.
We slide a square containing nine pixels on the mosaic image and fill the center blank cell of the square if the cell satisfies any of the following conditions:
\begin{description}
 \item[A] surrounded by five adjoining pixels painted in the same item color including two corners;
 \item[B] adjoining three connected pixels that include a corner with the same item color.
\end{description}
%
The first condition leads to canceling unnecessary concavities of the region painted in the item color.
The second one aims to sharpen a corner of a rectangle that may become an AOI.

Then, our technique sets temporary borders of AOIs, as shown in Figure \ref{fig_defineaoi}(c).
Starting from a pixel representing a corner of a region painted in the item color, our technique extends horizontal and vertical lines (the arrows in Figure \ref{fig_defineaoi}(c)) until finding the end of the region.
These lines are parts of the borders of an AOI.
Then, we add two lines to make a rectangle.
The borders surround the region completely when the region in the item color forms a rectangle as AOI 3 in Figure \ref{fig_defineaoi}(c).
Otherwise, the generated rectangle partially overlaps with the region as in case of AOI 1 and AOI 2.
Our technique applies this operation to all corner cells.
Finally, we arrange the borders of the temporary AOIs (Figure \ref{fig_defineaoi}(d)).
If two AOIs are adjoining or overlapping, we merge them or remove the smaller one.
We apply the merge of the two AOIs only when other AOIs will not be included in the new large AOI (purple rectangles in Figure \ref{fig_defineaoi}(d)).
There are two effects of this operation.
One is the reduction of overlapping AOIs.
The other is to merge multiple AOIs that substantially cover one common item.

Now, the technique can specify fine AOIs and display them upon a stimulus when the above processes are completed.
Then, we require users to set a hierarchy to the AOIs and set coarse AOIs which we mentioned in Figure \ref{fig_hierarchy}.
%
\begin{figure}[!h]
 \centering 
 \includegraphics[width=\columnwidth]{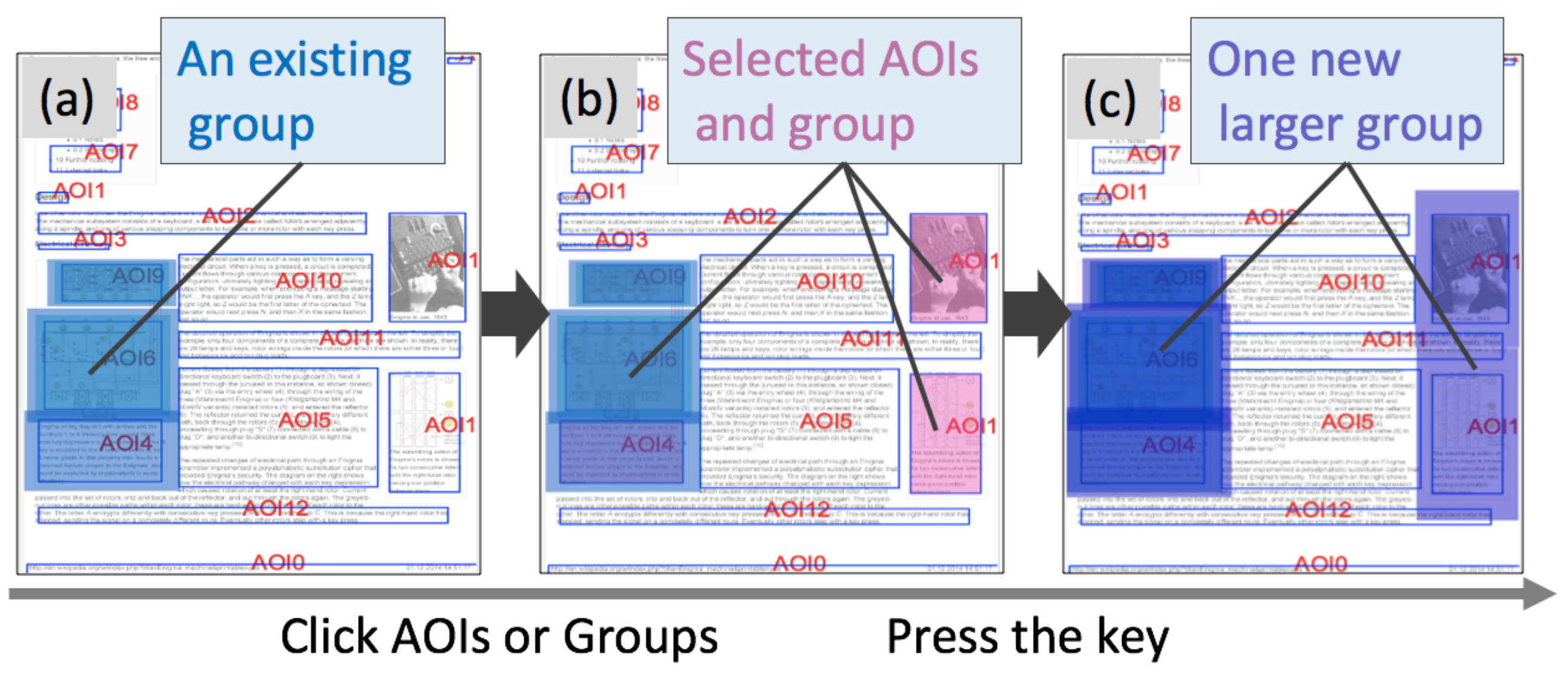}
 \caption{An example of making groups of AOIs.}
 \label{fig_makegroup}
\end{figure}
%
Specifically, users can define a larger AOI by making a group involving multiple AOIs.
Figure \ref{fig_makegroup} shows an example of this operation.
In the stimulus, the AOIs appear as rectangles outlined in blue without being filled, and groups are represented as filled rectangles, as illustrated in Figure \ref{fig_makegroup}(a).
Users can select AOIs or groups to place them in a new higher-level as one group by clicking AOIs that are drawn on a stimulus.
If the clicked AOI already belongs to a specific group, all other AOIs belonging to the same group also get selected.
Users can select several groups and make a larger group by this manipulation.
The selected items are painted in pink at this moment (Figure \ref{fig_makegroup}(b)), and a new group appears with the selected AOIs when a user presses a keyboard shortcut (Figure \ref{fig_makegroup}(c)).
The users can restart selecting AOIs to create the next group at this moment.
The generated group gets painted in sky blue, blue, or purple.
The painted region as one group covers a broader area than the exact size of an AOI and the existing groups.
This representation is helpful to check the current structure of the hierarchy.
Users can stop the process to set the hierarchy at any time.
Such manual operation can strongly reflect users' assumption on relationships between AOIs.
On later visualization step, the users can check how different the imagined relationships and actual AOI transitions.

We can describe the hierarchical AOIs using a tree diagram, as shown in Figure \ref{fig_hierarchy}.
The tree has as many leaves as the number of the AOIs unless they create groups.
A newly generated group gets inserted into the tree as a parent node of the selected AOIs and groups.
The tree gets deeper if users create groups by including particular AOIs many times.
The depth of the tree is not limited.
$L_k$ indicates the depth of hierarchy and fineness of AOIs.
As described in Section \ref{sec_force}, users can select which AOI definition to display by setting $L_k$ as a parameter.

\subsection{Generation of Strings and Pattern Extraction}
\label{sec_generation}

\begin{figure}[!h]
 \centering 
 \includegraphics[width=\columnwidth]{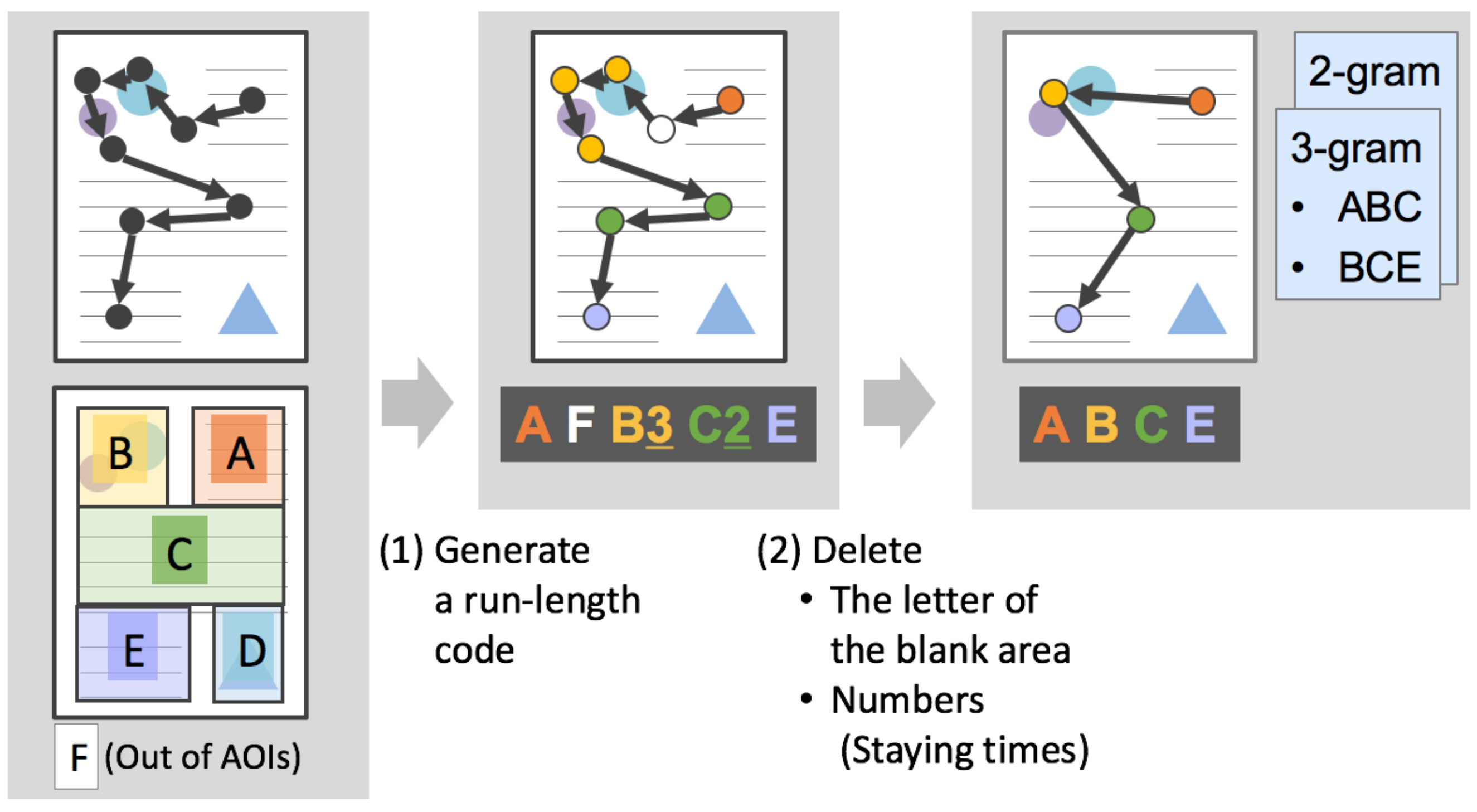}
 \caption{Generation of a run-length code and applying N-gram.}
 \label{fig_string}
\end{figure}

The target of our study is to extract major transition patterns through AOIs.
We apply N-grams to the characterized eye tracking scan-paths to extract the patterns.
This section describes how to characterize the scan-paths and apply N-grams.
The technique applies the following processes to find trends and patterns in eye-tracking data.
Figure \ref{fig_string} illustrates the steps.
First, we convert scan-paths to simple codes using the definition of AOIs (Figure \ref{fig_string}(1)).
A unique character is assigned to each AOI.
In addition, an unassigned character is used to label the blank area outside AOIs.
Using this allocation, the technique converts one scan-path $C_{i}$ into a string.
Each point $c_{ij}$ belongs to one AOI or the blank area.
We define that $c_{ij}$ changes to the character $c^{\prime}_{ij}$ and the string $C^{\prime}$ as follows.

\begin{equation}
  C^{\prime}_{i} = \{c^{\prime}_{i1} \; c^{\prime}_{i2} \; ... \; c^{\prime}_{in_{i}}\}
  \label{eq:string}
\end{equation}
%
Next, our technique applies run-length encoding to each $C^{\prime}_{ij}$.
This encoding extracts the continuous appearance of the same characters in $C^{\prime}_{i}$.
It then converts them to a pair of the character $m_{j}$ and the number of the continuous appearance $t_{j}$ .
We define the pair of the characters and the numerals as a unit $u_{j}$ .
%
\begin{equation}
	u_{j^{\prime}} =
	\begin{cases}
		m_{j^{\prime}} \;(t_{j^{\prime}} = 1)  \\
		m_{j^{\prime}}t_{j^{\prime}}\;\text{(otherwise)} \\
	\end{cases}
\label{eq:u}
\end{equation}
%
A run-length code is a sequence of this unit. $m_{j^{\prime}}$ denotes one AOI or the blank area that the participant looked at, and $t_{j^{\prime}}$ indicates the length of gazing at $m_{j^{\prime}}$.

The generated codes reflect the detailed AOIs.
We convert these codes corresponding to the coarse AOIs.
This process leads to the generation of multiple strings that represent rough movement to fine transitions using single eye tracking scan-path.
We define further rules to convert a character that means an AOI or a group to another character corresponding to a larger group using the hierarchical structure of the AOIs.
For example, when a group includes three AOIs called A, B, and C, one of the three characters gets selected to describe the larger group.
We use a character that corresponds to the largest AOI in the group.
If A is selected to describe the group, B and C in the generated strings change to A.
After repeating this conversion, only the character that means the blank area and another character for all items are finally used to compose strings.
%

Then, we apply a pattern extraction method using these codes.
N-grams are effective to find the order of looking at several particular AOIs.
We remove the characters representing the blank area and the numbers representing the staying time from the strings (Figure \ref{fig_string}(2)).
In addition, we can shorten the transitions by removing the units including a tiny number.
Such units denote that the participant just approached but not focused on the AOIs corresponding to the units.
The codes after the removal include only the characters corresponding to the AOIs; in other words, these codes clearly describe transitions among AOIs.
We apply N-grams to these characters and find patterns of transitions between the AOIs.
Roughly, we collect the following information:
%
\begin{itemize}
  \item Total number of each extracted pattern.
  \item Differences between the two participants.
\end{itemize}
%
First, we count the number of corresponding participants for each pattern.
This process leads to find a representative or exceptional behavior.
Second, we compare the differences in behaviors between arbitrary pairs of participants.
Specifically, we compare the kinds of included patterns and their appearance frequency.
At the same time, we calculate the cosine similarity between two strings for all pairs of participants.
The calculated values indicate the difference in behavior of two participants.

\subsection{Visualization of Extracted Patterns}
\label{sec_visualization}

\begin{figure}[!h]
 \centering 
 \includegraphics[width=\columnwidth]{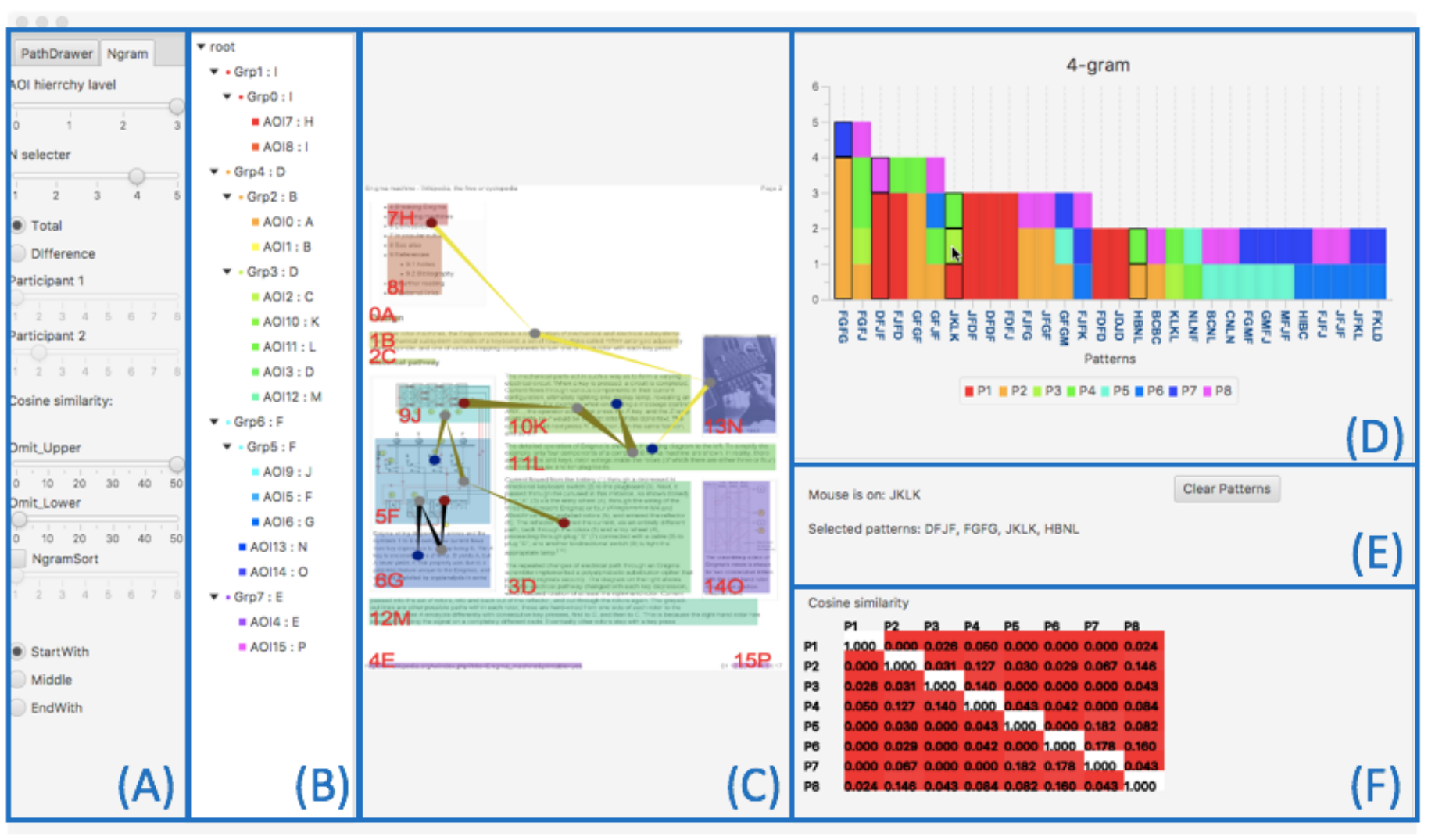}
 \caption{The user interface for visualizing eye-tracking data.}
 \label{fig_gui}
\end{figure}

Finally, we represent the results of N-grams using two visualization components.
Figure \ref{fig_gui} shows our user interface.

%

Users can open datasets and set parameters for visualization in the ``Control panel'' shown in Figure \ref{fig_gui}(A), and have an overview of the hierarchical structure of AOIs using ``Tree view'' shown in Figure \ref{fig_gui}(B).
``AOI view'' shown in Figure \ref{fig_gui}(C) displays the stimulus and specific AOIs.
Users can visualize specific paths of extracted patterns here.
``AOI view'' is also useful for defining and grouping AOIs (see Section \ref{sec_definition}).
``Bar chart view'' shown in Figure \ref{fig_gui}(D) is another visualization component to show the results of N-grams.
Moreover, users can observe the patterns visualized in the AOI view using ``N-gram view'' shown in Figure \ref{fig_gui}(E).
Figure \ref{fig_gui}(F) shows ``Matrix view,'' which represents the cosine similarity values 
described in Section \ref{sec_generation}, 
highlighting dissimilar pairs of participants in red.
In the following sections, we describe the details of the two main visualization components.

\subsubsection{Stacked Bar Chart for Showing Results of N-grams}
\label{sec_stacked}

We use ``Bar chart view'' to show results of N-grams.
Users can arrange the following parameters.

\begin{itemize}
  \item $k$ as level of AOIs.
  \item $N$ as the length of extracted patterns.
  \item ``Total'' or ``Difference''.
  \item Two participants to compare. (Only valid when selecting ``Difference''.)
  \end{itemize}

\begin{figure}[!h]
 \centering 
 \includegraphics[width=\columnwidth]{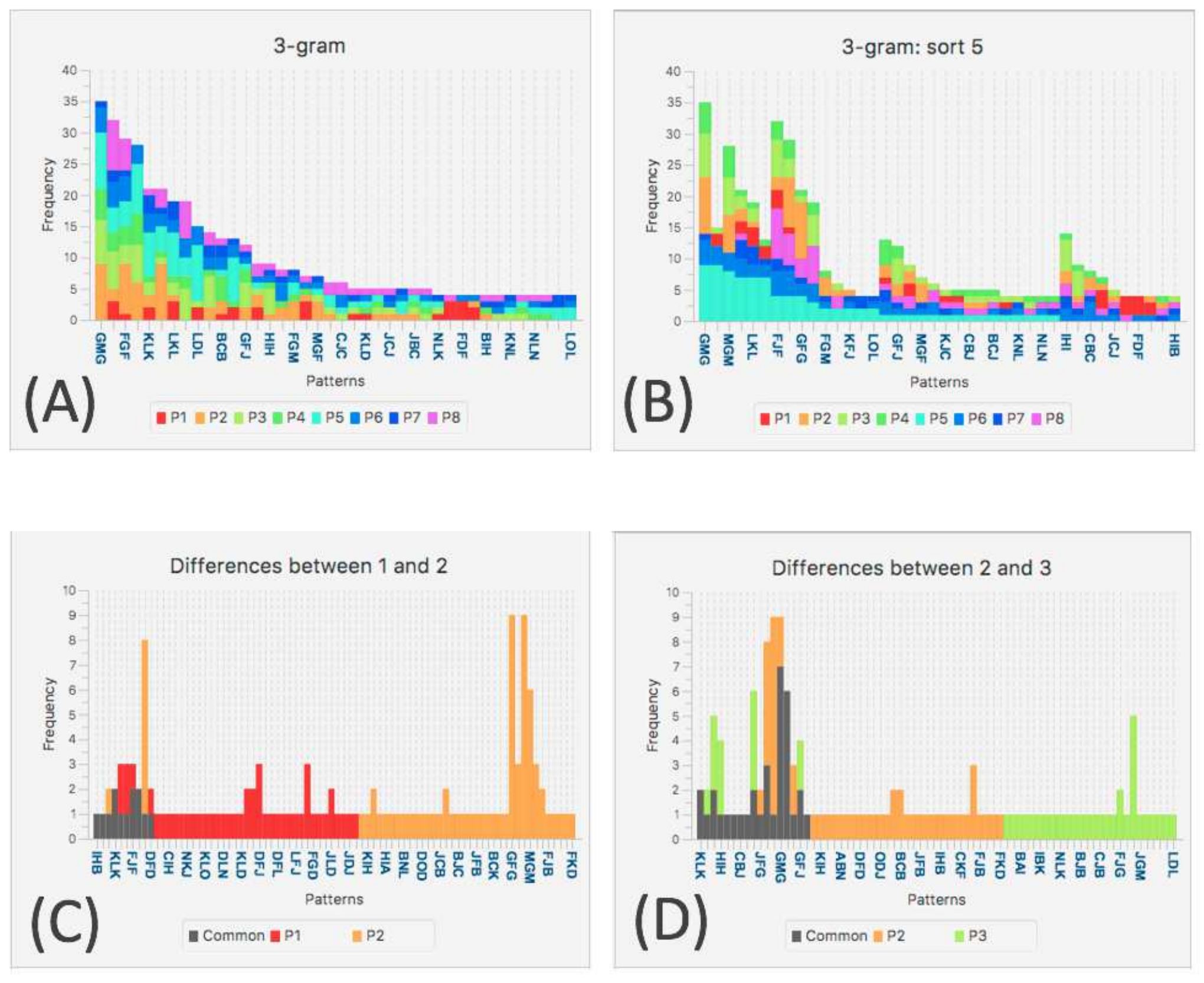}
 \caption{Examples of ``Bar chart view'' for displaying the results of N-grams.}
 \label{fig_ngramchart}
\end{figure}

$k$ means the fineness of AOIs.
The higher $k$ is, the finer AOIs get.
The highest $k$ is the same as the depth of the hierarchy of AOIs, and in this case, $L_k$ means detailed AOIs without any groups.
$N$ means a length of patterns to extract.
Users can select showing statistics of patterns or differences between two participants.
When showing differences, users can select two participants to be compared.
Figure \ref{fig_ngramchart} shows four examples of ``Bar chart view''.
\vspace{-0.5\baselineskip}
 \vspace{-0.5\baselineskip}
In ``Bar chart view,'' specific patterns extracted by N-grams are assigned to the X axis, and values of Y coordinate describe the frequency of their appearances.
Users can select to show only particular patterns that appeared more (or less) than a threshold.
Colors correspond to participants in the chart.
In the total visualization, firstly patterns are lined up in descending order, as shown in Figure \ref{fig_ngramchart} (A).
We can observe representative or unusual behaviors with this visualization.
Users can apply the sort process focusing on one selected participant, as shown in Figure \ref{fig_ngramchart} (B).
At this moment, the focused participants appear at the lowest part as sky blue bars, and other bars are stacked on them.
This sort is helpful to find which patterns the participant belonged to or not and briefly compare to other participants.
If the users selected the difference visualization shown in Figures \ref{fig_ngramchart} (C) and (D), bars are classified into three types.
Gray bars in the left area depicts the common behavior in both of the selected participants.
If the top of a bar is colored in another color corresponding to one participant, the colored bar depicts differences in frequency.
In contrast, bars in different colors in the center and right area depict unique behaviors.

\subsubsection{A Force-Directed Graph Layout to Describe Patterns}
\label{sec_force}

%
The other visualization is a force-directed graph layout on the stimulus and AOIs for showing specific movements in ``AOI view''.

First, our implementation applies the alpha-compositing to the stimulus whiter than the original picture to keep the visibility of the graph overlaid on the picture.
The second step is drawing AOIs on the stimulus.
Each AOI is represented as a colored translucent rectangle.
The ID of the AOI and the character used as codes in the paths are indicated in the lower left part of the AOI.
The colors of the AOIs depict the structure of the groups corresponding to the depth of the hierarchy that the user selects to display.
The common color is assigned to the AOIs belonging to the same group when the user selects the coarse division of AOIs.
In contrast, if the user chooses to draw all individual AOIs, a unique color is assigned to each of the AOIs.
Similar colors are assigned to the AOIs that belong to a common group so that we can indicate their relationship even while applying the latter case.
%
%
\begin{figure}[!h]
 \centering 
 \includegraphics[width=\columnwidth]{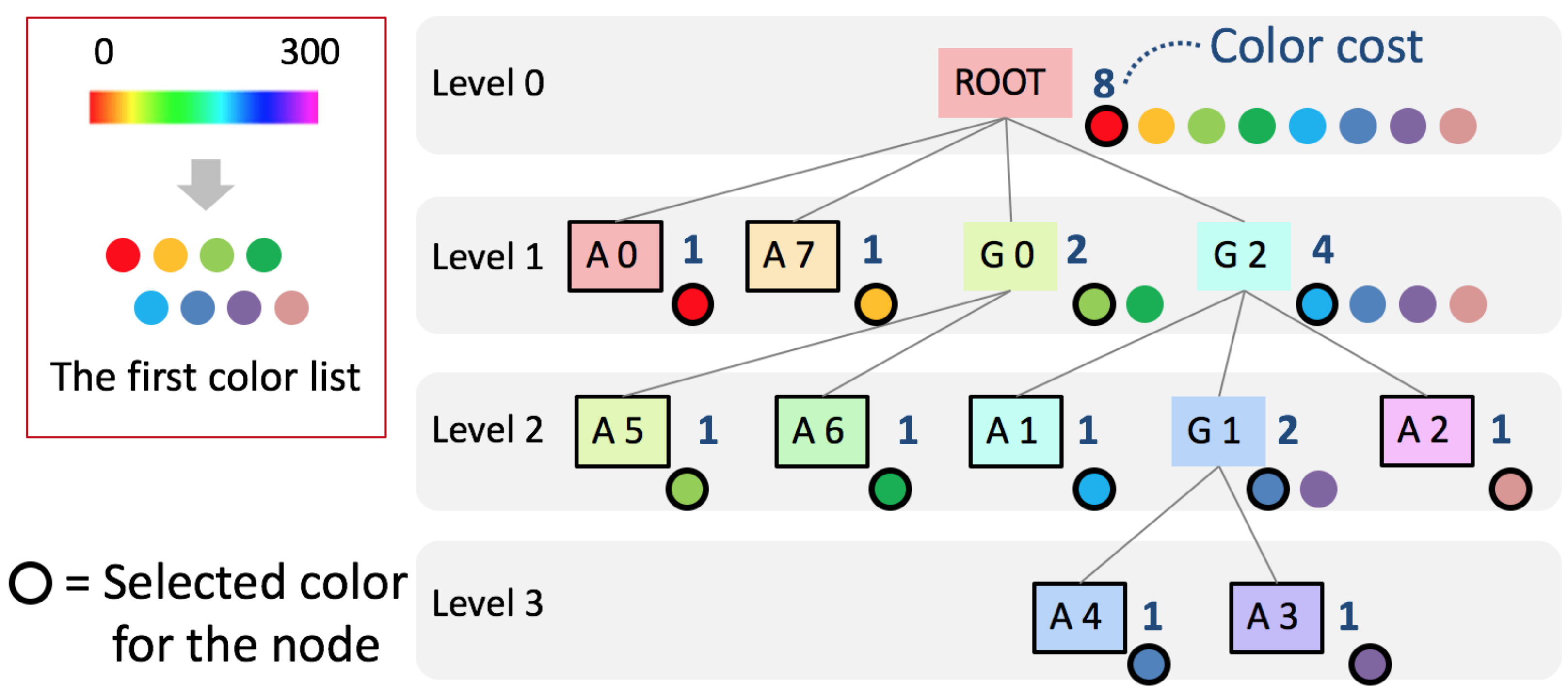}
 \caption{Distribution of colors to each AOI and group.}
 \label{fig_colorcost}
\end{figure}
%
The detailed processing flow of assigning colors to AOIs is as follows.
Figure \ref{fig_colorcost} is an example of this process.
Note that A encodes an AOI and G a group, in the tree diagram.
The first process is to calculate the costs of colors for each node.
A cost of a single AOI, which is a leaf of the tree, is one.
Costs of parent nodes are the sum of costs of their child nodes.
Following this rule, the cost of the root node finally gets the same value as the number of AOIs.
Then, the colors of all AOIs and groups are selected based on these costs.
We create a list of hue values that will be assigned to the AOIs and the groups.
Our implementation sets the minimum value of the hue as 0 and the maximum value as 300.
From this range, as many colors as the number of AOIs are selected at regular intervals.
One of the colors in the color list is assigned to each node and the divided lists are transferred to child nodes starting from the root node.
A child node receives as many colors from its parent node as its color cost.
Consequently, each leaf node receives a single color.
This algorithm is effective making the AOIs that belong to a common group have similar colors.
Furthermore, even if the user switches the level of the hierarchy to draw, each AOI does not change the color sharply.
For example, if the user selects $L_2$ in Figure \ref{fig_colorcost}, AOIs except for A4 and A3 have unique colors.
A4 and A3 get the same blue color as G1.
This color is the same as that of A4 and similar to the purple color of A3 in $L_3$

\begin{figure}[!h]
 \centering 
 \includegraphics[width=\columnwidth]{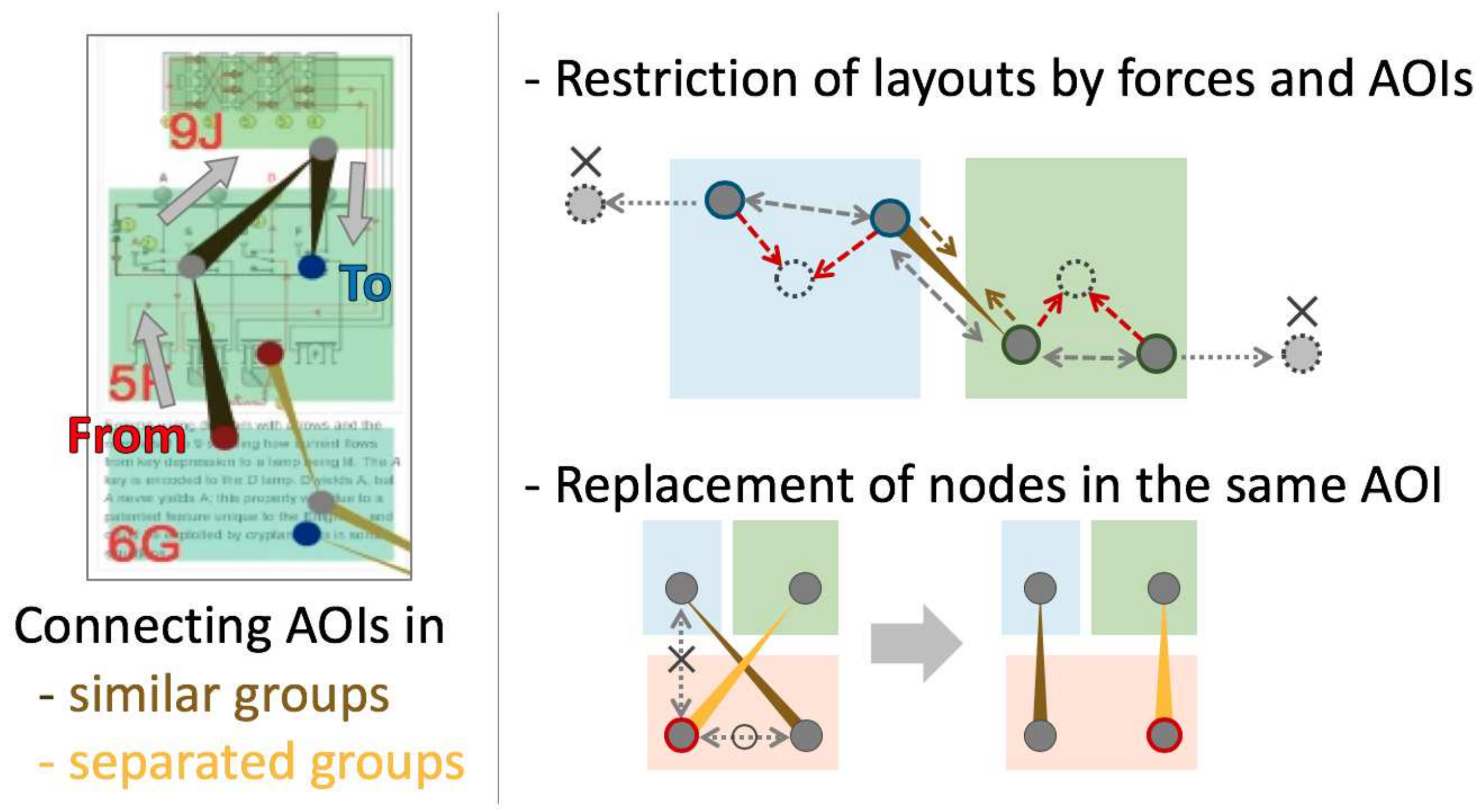}
 \caption{Layout of nodes and edges.}
 \label{fig_graph}
\end{figure}

Then, we draw a force-directed graph on the AOIs.
First, we require users to select patterns to visualize by clicking bars in the ``Bar chart view'' or clicking an AOI on the stimulus.
When the users choose an AOI, they can additionally select the following three types of patterns.
\begin{itemize}
\item Starts from the AOI
\item Passes over the AOI
 \item Arrives at the AOI
\end{itemize}
Figure \ref{fig_graph} shows an example of the graph and the rules of the layout.
Nodes appear in corresponding AOIs, and edges show transitions between two AOIs.
Red nodes depict start-points while blue ones depict end-points of the extracted patterns.
When a user opts to display hierarchical groups, we selection the largest AOI in each visible group and position the graph node at the center of the AOI.
Thus, nodes and edges do not always indicate the exact positions if the group includes multiple AOIs.
Edges are drawn with arrows to represent directions.
When a user moves a cursor on a particular bar in ``Bar chart view,'' the corresponding edges get thick.
The color of each edge depicts the strength of relations between AOIs included in one pattern.
On the other hand, edges among nodes in different groups get vivid yellow.
This can help find relationships between AOIs which users did not expect.
These nodes and edges follow the force-directed algorithm.
Moreover, we added further rules to reflect the layout of AOIs.
First, we added another force that attracts nodes to the center of each node to prevent nodes to extremely close to the border of the AOI.
Second, we apply to swap nodes to remove overlapping of edges.
We check crossing edges and select two nodes from each edge which are in the same AOI.
This layout can clearly show multiple transitions among AOIs in exchange for showing exact positions in an AOI.

\section{Experiment}
\label{sec_experiment}

\begin{figure}[!h]
\centering 
\includegraphics[width=\columnwidth]{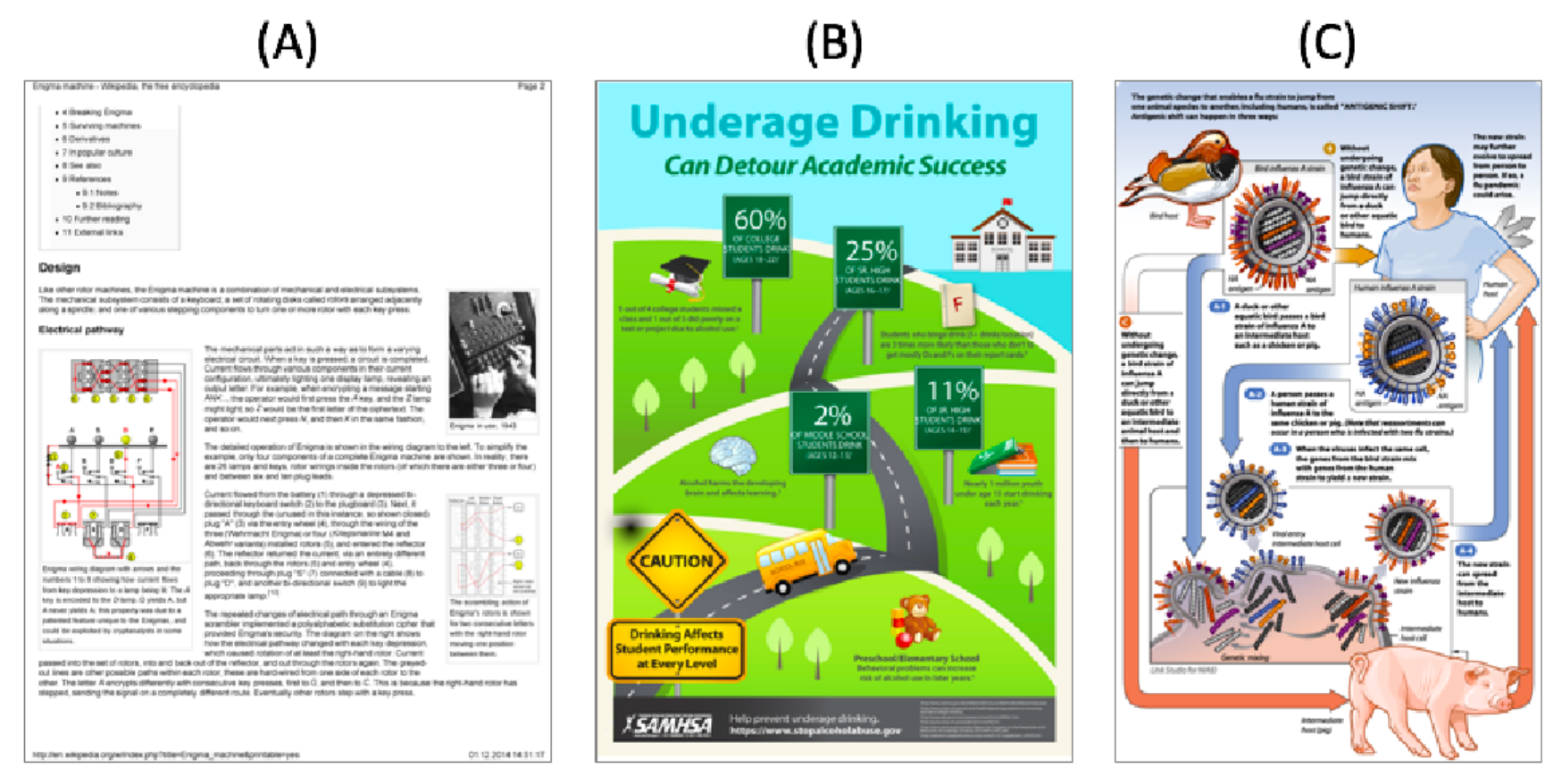}
 \caption{Three static stimuli. (A) A page of Wikipedia. (B) A poster on drinking for young people. (C) A description of antigenic shift.}
 \label{fig_stimuli}
\end{figure}

We present the results of applying our technique to three stimuli shown in
Figure \ref{fig_stimuli}.
(A) is a Web page of Wikipedia that describes ENIGMA.
(B) \footnote{\url{https://alcoholanddemocracy.wordpress.com/2017/04/10/155/}, Center for Substance Abuse Treatment. Detoxification and Substance Abuse Treatment. Treatment Improvement Protocol (TIP) Series 45. HHS Publication No. (SMA) 06-4131. Rockville, MD: Substance Abuse and Mental Health Services Administration, 2006.}
includes multiple pictures to show the ratio of young people who have drunk and the influence of alcohol.
(C) \footnote{\url{https://en.wikipedia.org/wiki/File:AntigenicShift_HiRes.svg}}
shows three types of mechanism of antigenic shift.
We asked eight participants to observe each stimulus for 90 seconds and collected 24 scan-paths totally.
We introduce the results in the following sections.

\subsection{Case A : A Page of Wikipedia}
\label{sec_caseA}

\begin{figure}[!h]
 \centering 
 \includegraphics[width=\columnwidth]{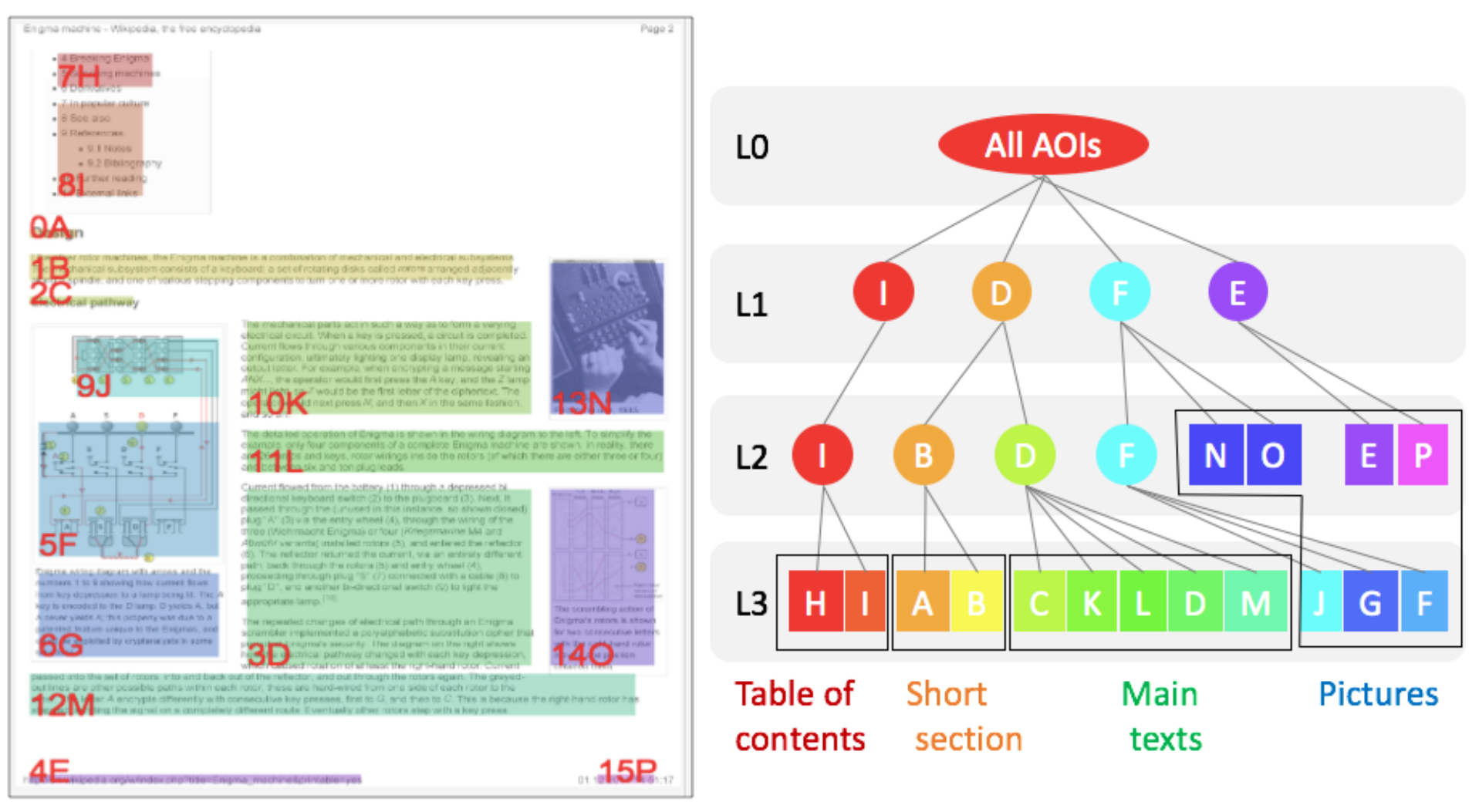}
 \caption{AOIs on the stimulus (A).}
 \label{fig_aoienigma}
\end{figure}

Figure \ref{fig_aoienigma} shows the hierarchical AOIs on the stimulus (A).
We roughly classified the contents to (1) table of contents, (2) the short section, (3) main texts, and (4) figures and the footer, as enclosed by black borders in this figure.
Then, we applied an automatic definition of fine AOIs.
The depth of the hierarchy structure is four, which corresponds to the fineness of the AOIs.


\begin{figure}[!h]
 \centering 
 \includegraphics[width=\columnwidth]{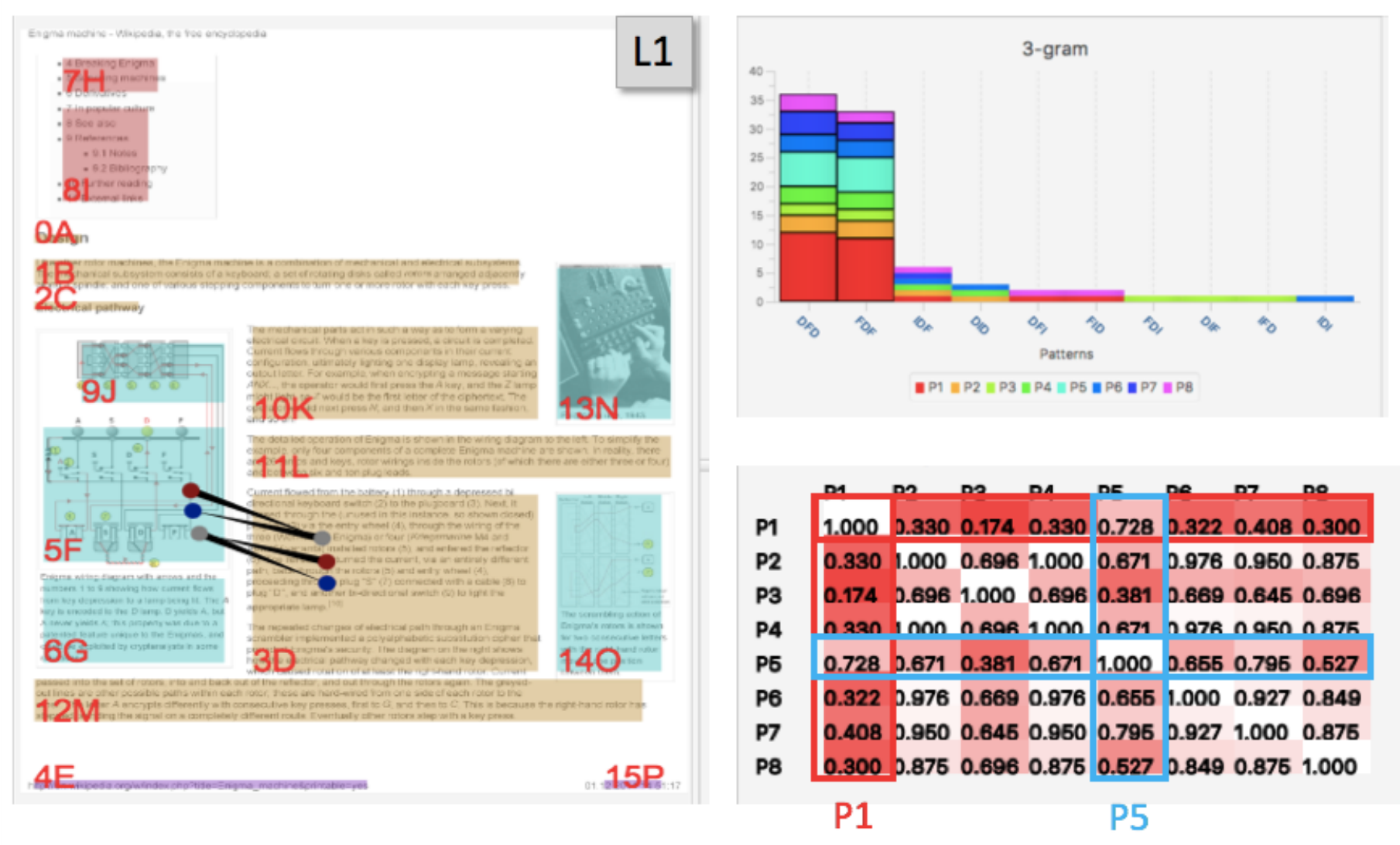}
 \caption{Brief transitions among text areas and picture areas.}
 \label{fig_enigma0}
\end{figure}

Figure \ref{fig_enigma0} shows frequencies of patterns extracted by 2-gram on the coarse AOIs ($L_1$).
The two very long bars in the ``Bar chart view'' correspond to movements between orange texts and blue pictures.
Especially, $P_1$ and $P_5$ had more transitions than other participants, which is also described in ``Matrix view''.

\begin{figure}[!h]
 \centering 
 \includegraphics[width=\columnwidth]{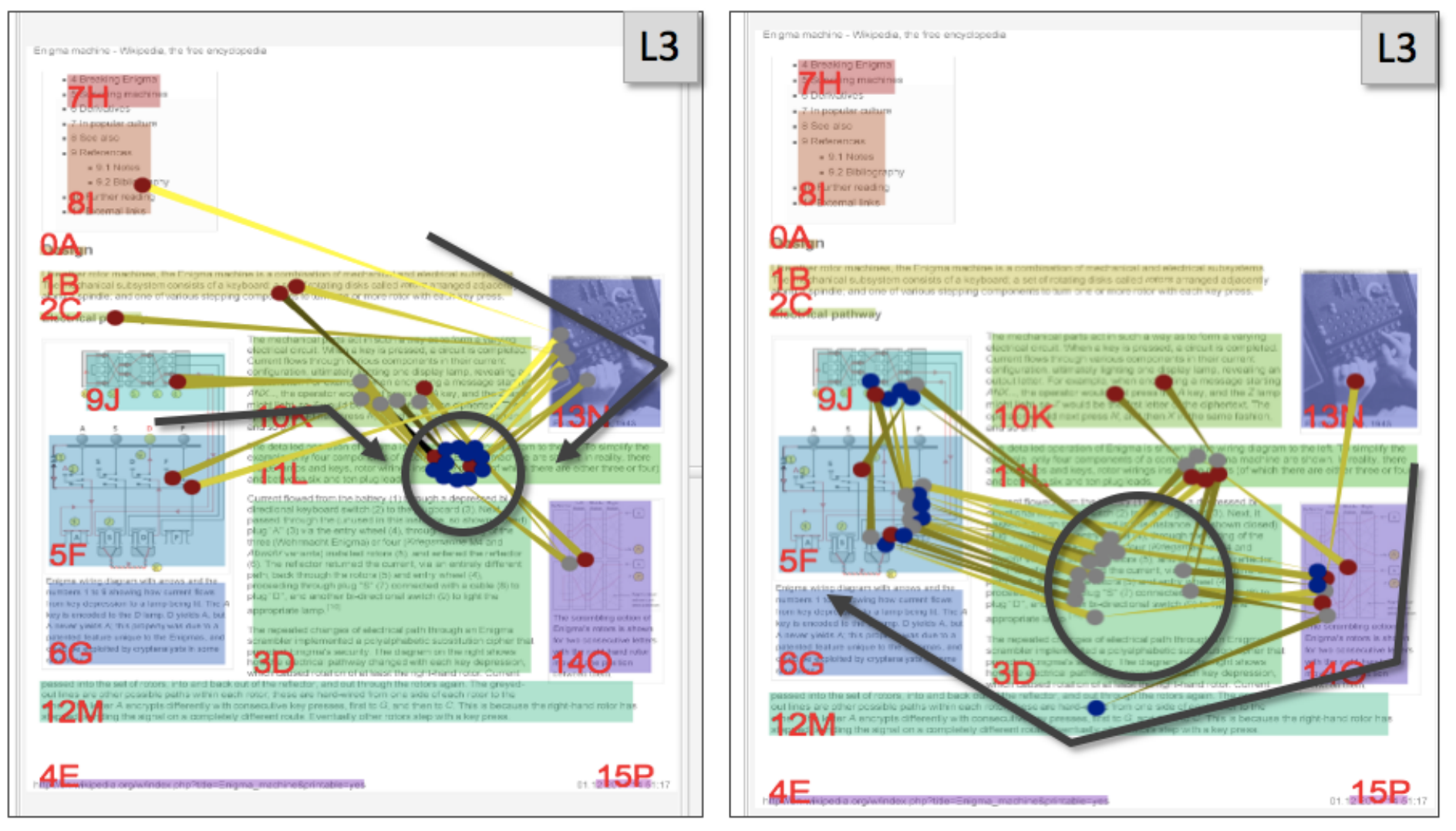}
 \caption{
Transitions on text paragraphs. (left) Patterns extracted by 3-grams that reached the circled paragraph. (right) Patterns extracted by 4-grams that passed the circled paragraph.
 }
 \label{fig_enigma1}
\end{figure}

Figure \ref{fig_enigma1} introduces transitions between multiple AOIs on the fine AOIs ($L_3$) establishing patterns that terminate on the circled paragraph.
We can mainly find many gray nodes, which shows movements following the regular order for reading.
We can also find that several patterns came from the upper right picture, where several other gray nodes are on the picture.
Notice that a vivid yellow edge starts from the table of contents on the upper left side.
This edge depicts a pattern that supposedly passed weakly related AOIs, specifically the table of contents, the right picture, and the selected paragraph.
The picture seems to collected attentions as well as the first green paragraph, even though the picture is far from the table of contents on the upper left of the stimulus.
Figure \ref{fig_enigma1}(right) shows the result of 4-gram highlighting patterns which passed the center large text area.
Many of red nodes are on upper or right AOIs, while blue nodes are noticeable on the left picture.
Also, there are no edges that straightly reached to the next paragraph.
The results depict the participants turned to pictures during reading the main texts.

\subsection{Case B : An Advertisement Without a Clear Direction of Reading}
\label{sec_caseB}

\begin{figure}[!h]
 \centering 
  \includegraphics[width=6cm]{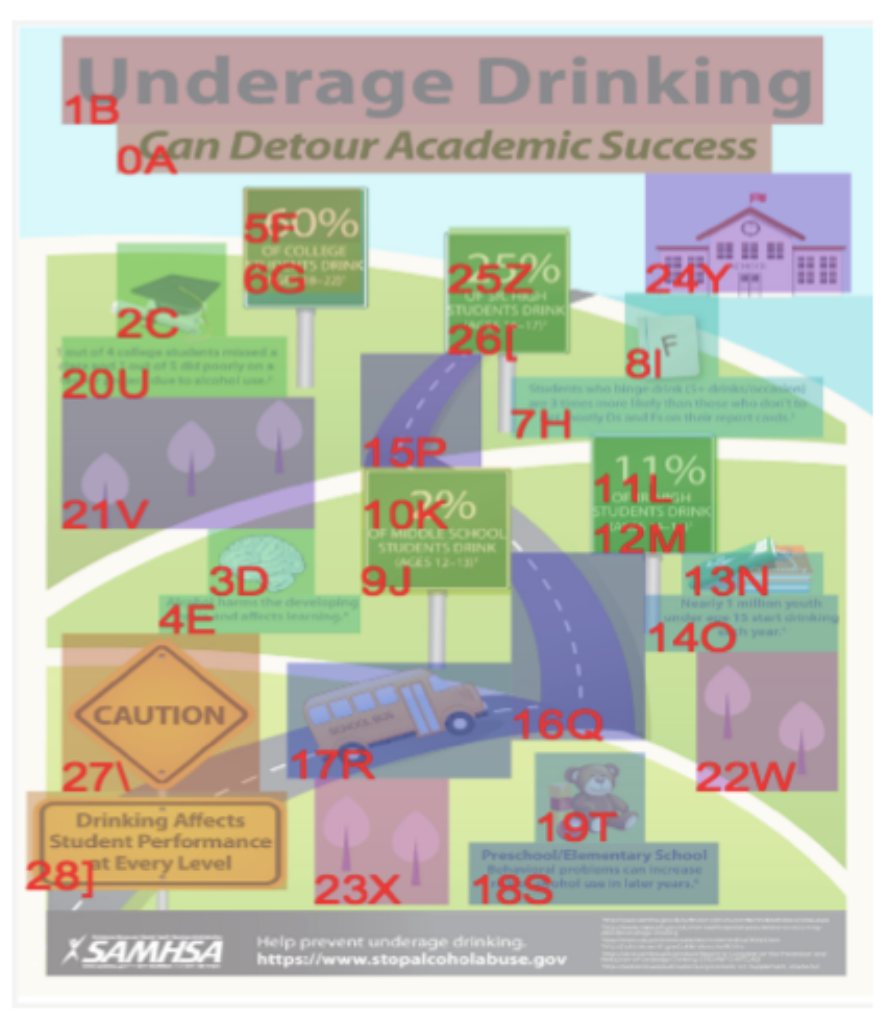}
 \includegraphics[width=\columnwidth]{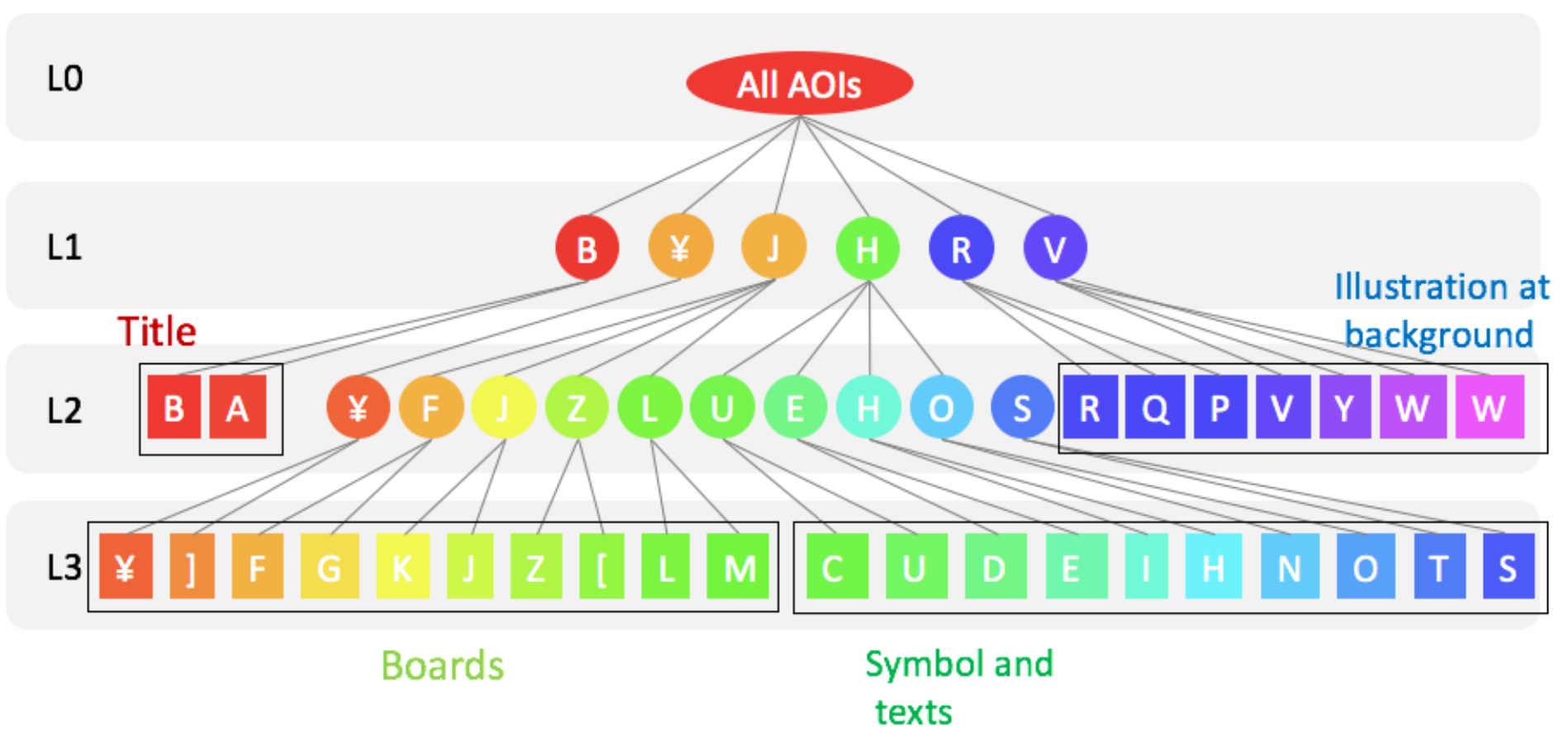}
 \caption{AOIs on the stimulus (B).}
 \label{fig_aoiuad}
\end{figure}

Figure \ref{fig_aoiuad} shows the hierarchical AOIs on the stimulus (B).
We manually set both of the AOIs and the hierarchy.
The figure's design is different from the other stimuli that there are several regions that consist of a symbol and text description below.
The green signs highlight percentages of people who are under age defined in the lower texts but already have drunken alcohol before.
Similarly, several icons on the green grass match the contents of the neighboring texts.
We firstly separated such pairs into two separate AOIs and rejoined them into groups to preserve the semantics.


\begin{figure}[!h]
 \centering 
 \includegraphics[width=\columnwidth]{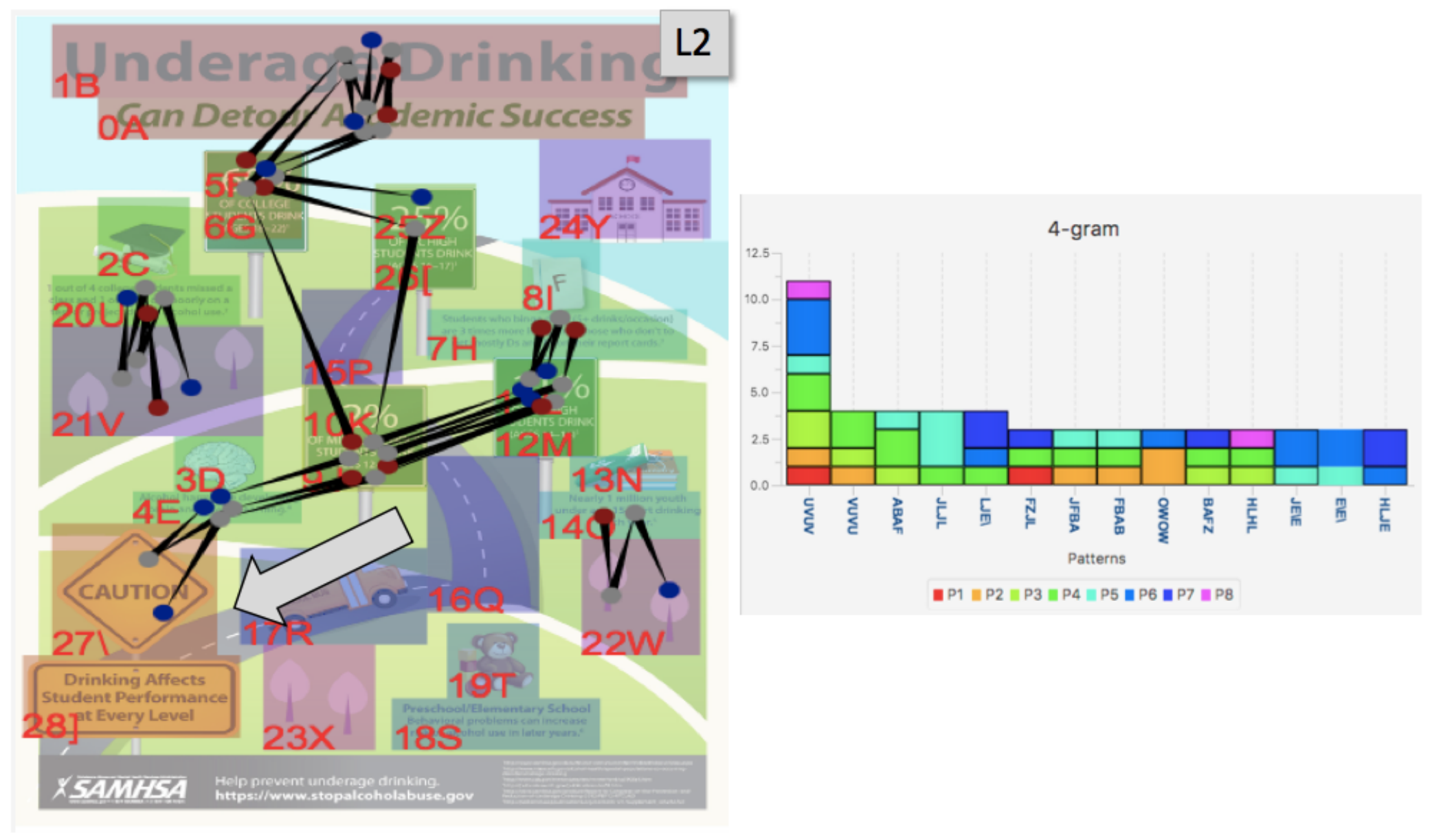}
 \caption{Frequent patterns extracted by 4-grams on the stimulus (B).}
 \label{fig_uad0}
\end{figure}

Figure \ref{fig_uad0} shows the frequent patterns extracted by 4-grams on the AOIs. 
We can find several patterns that start from the upper right picture or a board, pass the next board, and finally reach the yellow board on the lower left side of the stimulus.
This visualization indicates that many participants observed the stimulus from top to bottom.
Especially, almost all participants had as most frequent pattern connecting upper left AOIs U and V.
This is, however, the opposite of the supposed order.
If we start on the bus and follow the road, we can trace changes in ages.
Furthermore, the yellow boards may be read late even though they contain the summary message that ``Drinking Affect Student Performance at Every Level'' with vivid color.
The design does not seem effective to aid the order of reading to follow the content.

\begin{figure}[!h]
 \centering 
\includegraphics[width=\columnwidth]{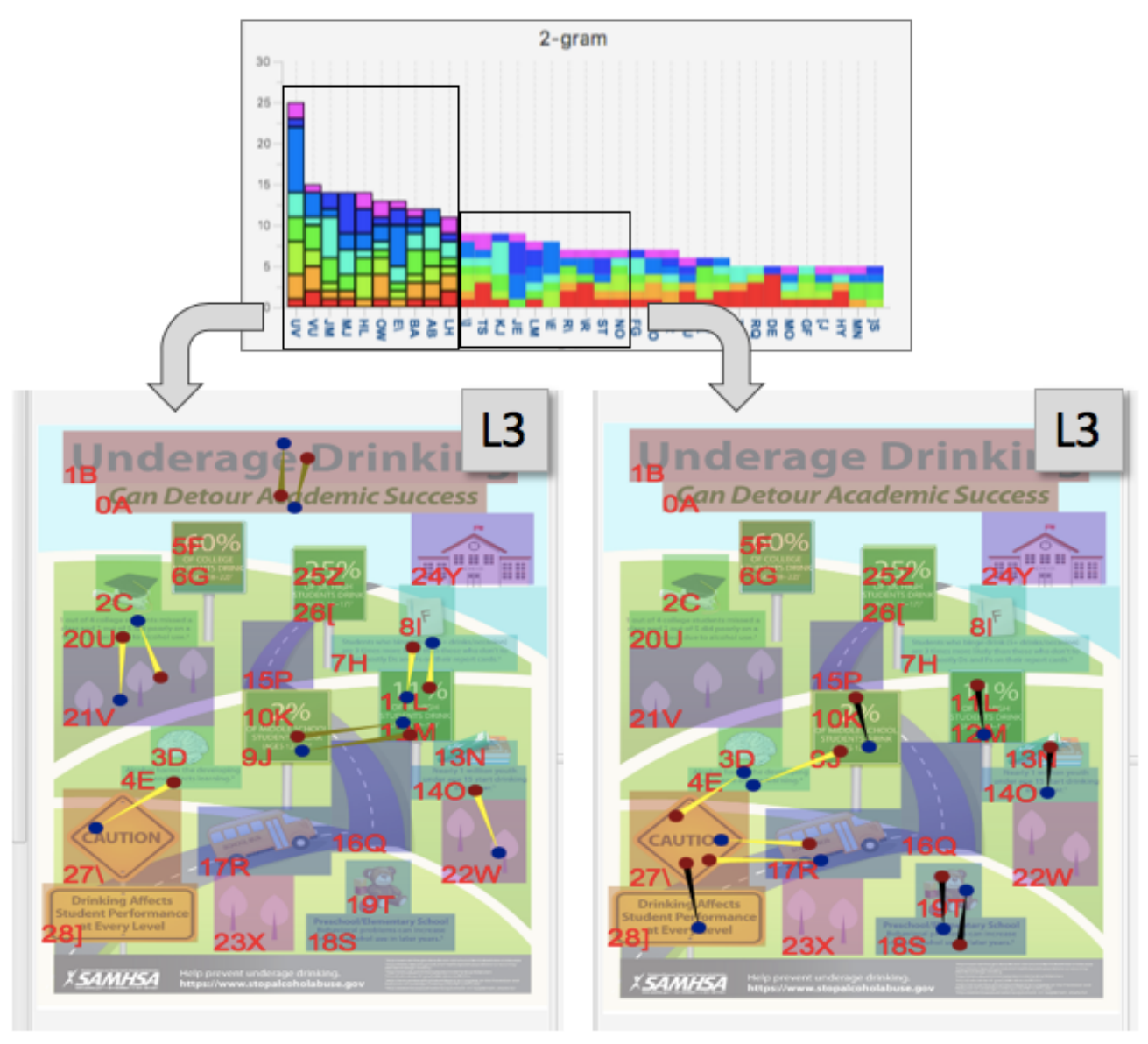}
 \caption{Comparison of frequent patterns extracted by 2-grams. (left) Patterns for the top 10. (right) Patterns from 11th place to 20th.}
 \label{fig_uad1}
\end{figure}
%
The bars in the left side of ``Bar chart view'' shown in Figure \ref{fig_uad1} correspond to the top ten patterns.
Some of these patterns connect two different objects, such as two green boards on center.
Especially, there are six yellow edges connecting two AOIs that do not seem to be directly related.
In contrast, there are multiple black edges that connect a symbol and its description, in the 11th to 20th patterns corresponding to the next ten bars in ``Bar chart view''. 
In other words, such connections are not the most representative movements.
The participants might tend to access detailed texts rather than the symbols.
On the whole, the participants did not follow the indicated order of reading.
They started checking contents on the upper side and moved to the lower side, corresponding to the regular reading order of texts.
Furthermore, they sometimes traced only texts without focusing on the symbols with the texts.
As for the series of some objects, many participants traced four green boards on the upper side continuously.
However, five pairs of an upper picture and a lower text were not strongly connected.

\begin{figure}[!h]
 \centering 
\includegraphics[width=\columnwidth]{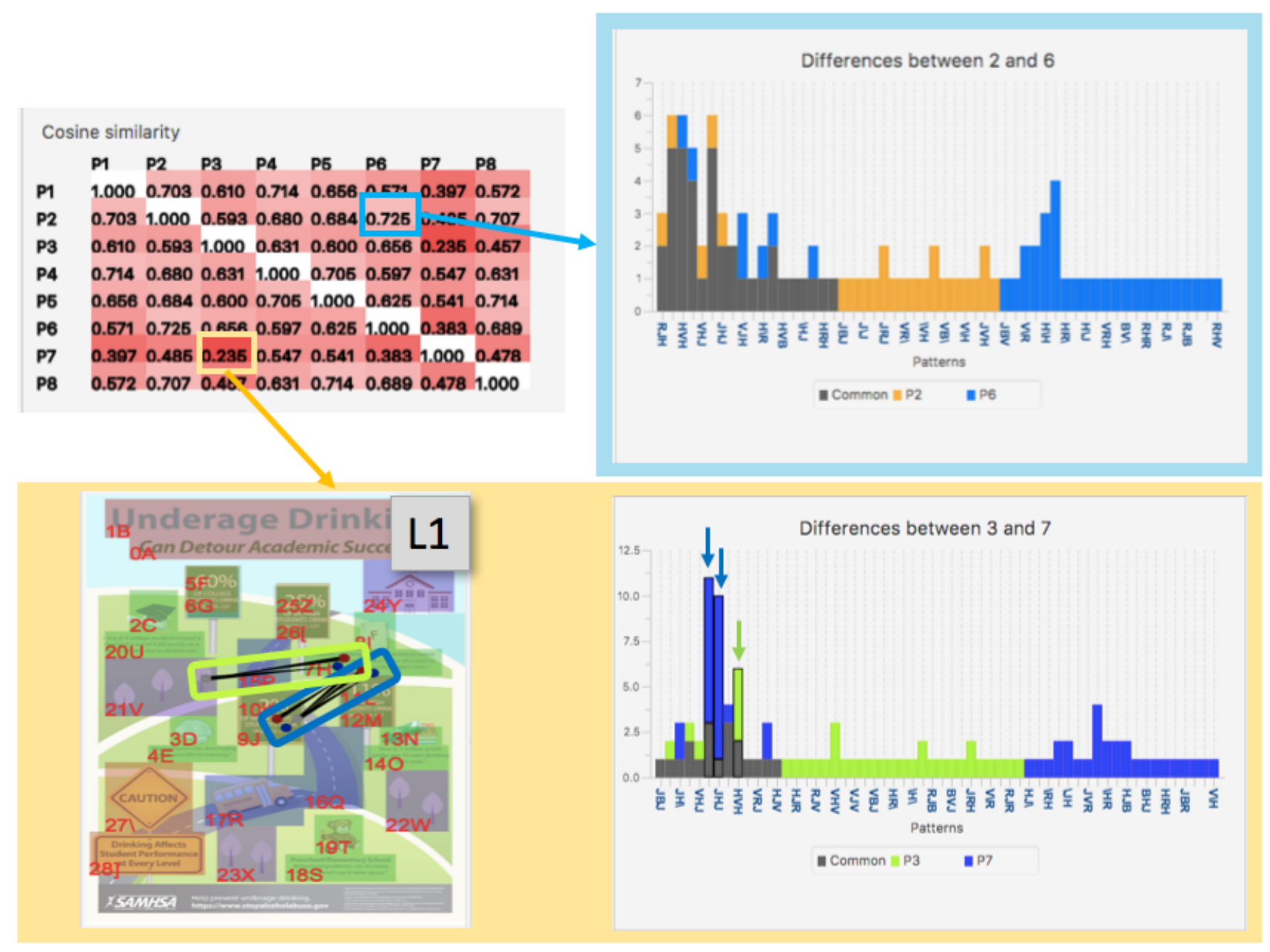}
 \caption{Comparison of similar and non-similar participants.}
 \label{fig_uad2}
\end{figure}
%
Figure \ref{fig_uad2} shows a comparison of several participants using the difference visualization on the coarse AOIs $L_1$.
We calculated the cosine similarity and selected the most similar / non-similar participants.
The upper sky-blue part contains a visualization of differences in similar participants $P_2$ and $P_6$.
The cosine similarity is 0.725.
We find that almost one-third of patterns are common and there are little differences in frequency.
In contrast, the bar chart in the lower yellow part depicts differences between non-similar participants $P_3$ and $P_7$.
The cosine similarity is 0.235.
We selected three particularly long bars to visualize.
Two blue bars correspond to transitions between boards and texts under symbols.
This shows that $P_7$ had much more transitions than $P_3$.
The green bar corresponds to the transition between texts and illustrations on the background.
$P_3$ had more of this transition than $P_7$. 
This result indicates that $P_7$ paid more attention than $P_3$ to specific information in the poster.

\subsection{Case C : A Poster with Strongly Displayed Directions of Reading}
\label{sec_caseC}

\begin{figure}[!h]
 \centering 
   \includegraphics[width=6cm]{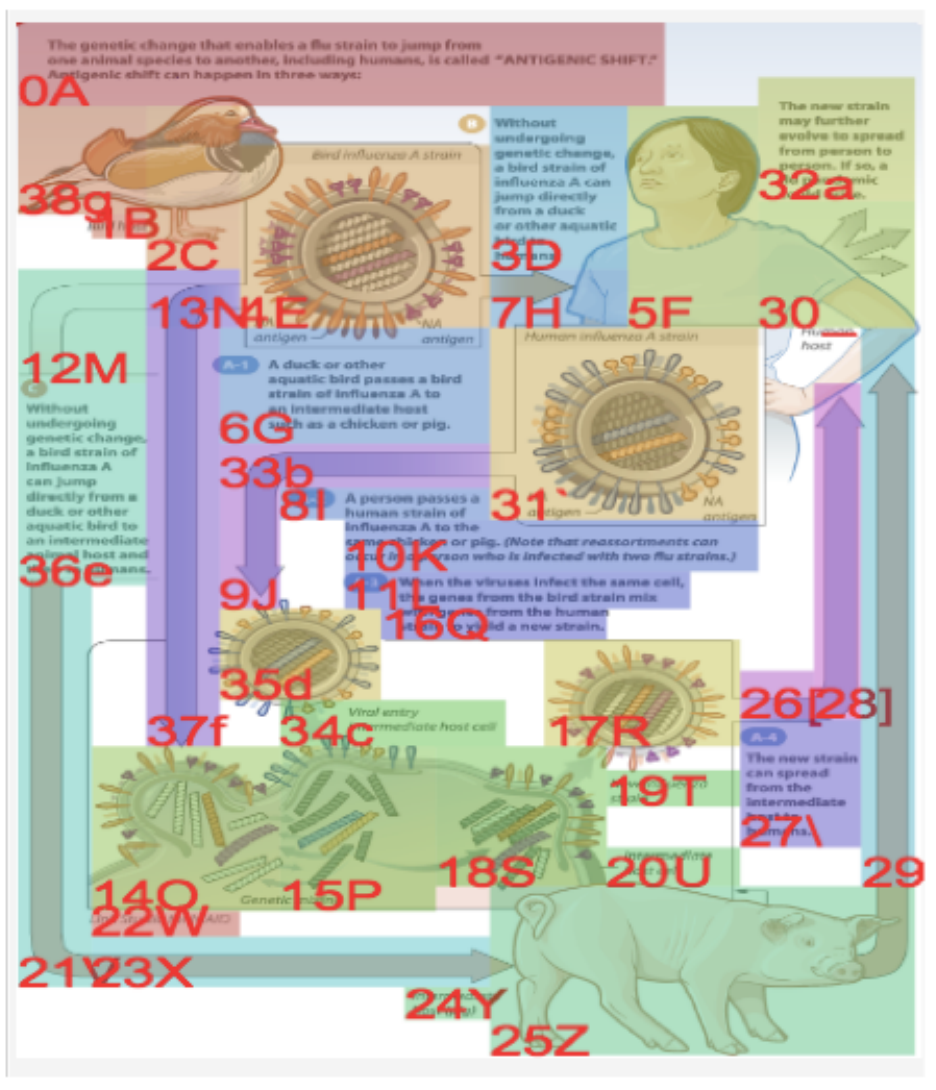}
  \includegraphics[width=\columnwidth]{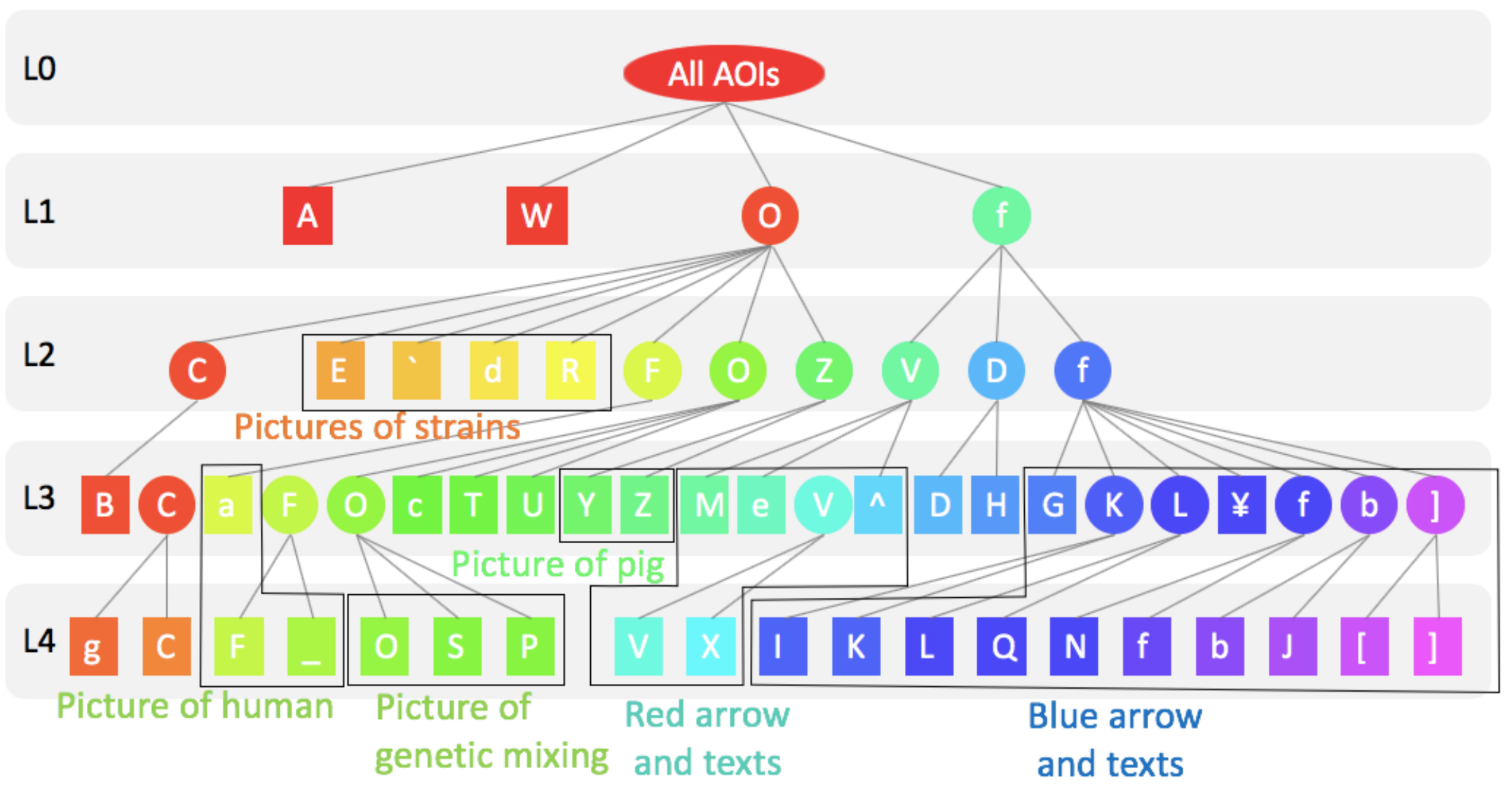}
 \caption{AOIs on the stimulus (C).}
 \label{fig_aoiantigenic}
\end{figure}

Figure \ref{fig_aoiantigenic} illustrates the AOIs of the stimulus (C).
We manually set the AOIs and the hierarchy.
We classified contents to ``blue arrow route,'' ``red arrow route,'' ``yellow arrow route,'' and the other pictures.

\begin{figure}[!h]
 \centering 
  \includegraphics[width=\columnwidth]{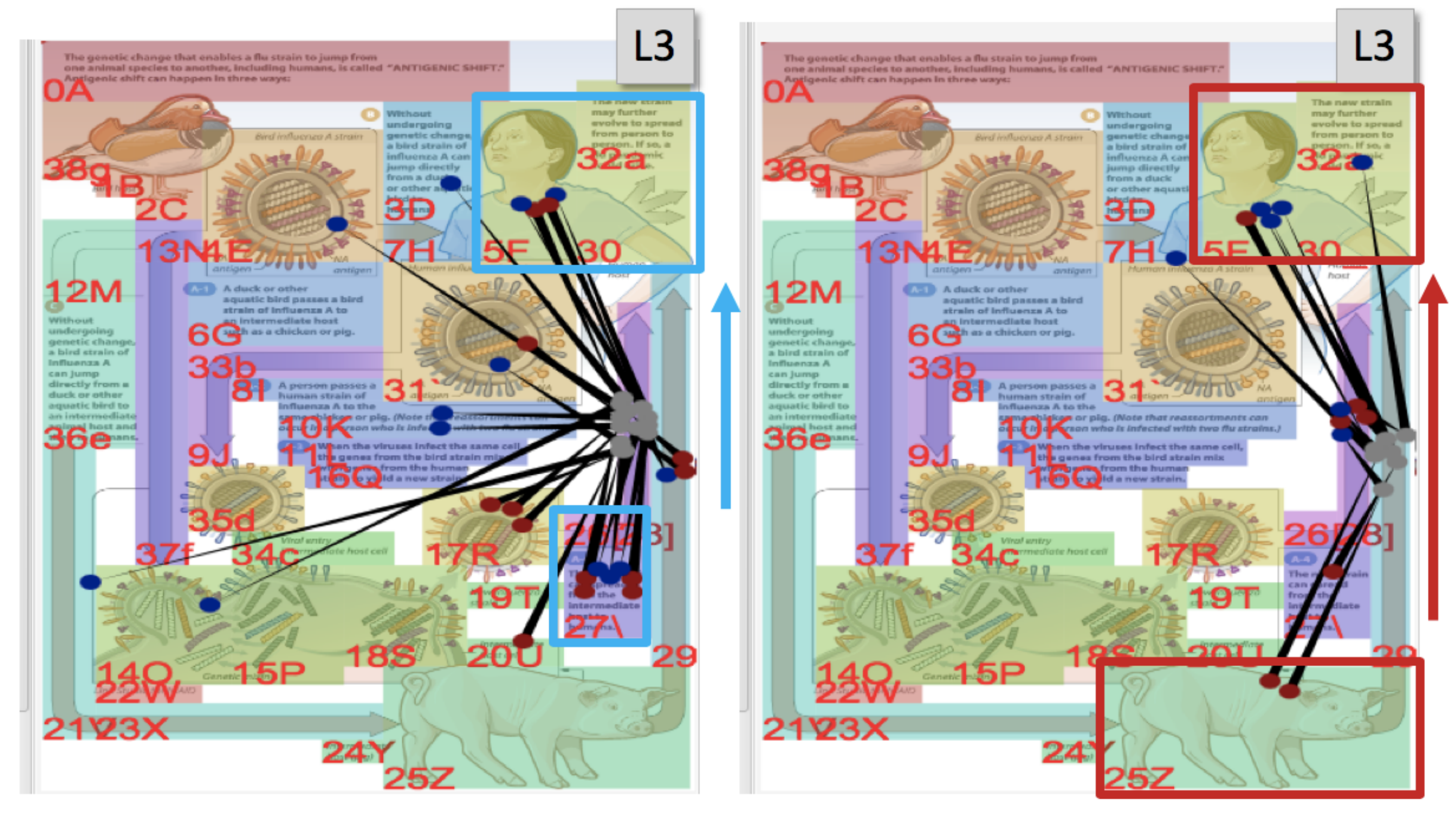}
 \caption{
 Patterns that follow the two straight arrows on the right side extracted by 3-grams.
 }
 \label{fig_antigenic0}
\end{figure}

We focused on the analysis of the effects of the arrows on this stimulus. 
Figure \ref{fig_antigenic0} shows the patterns passed along the arrows on the right side of the stimulus.
Along the blue arrow in the left picture, there are gathered edges between the text and human.
These edges indicate that participants followed this direction.
Some edges went to left texts before reaching the upper picture; however, the texts are in the same blue route and related contents and not mean deviation.
However, no movements clearly deviated on the right figure.
These arrows in the stimulus are easy to follow because they are at the edge of the stimulus and separated from other contents.
We suppose that is why the participants followed the arrows.

\begin{figure}[!h]
 \centering 
  \includegraphics[width=\columnwidth]{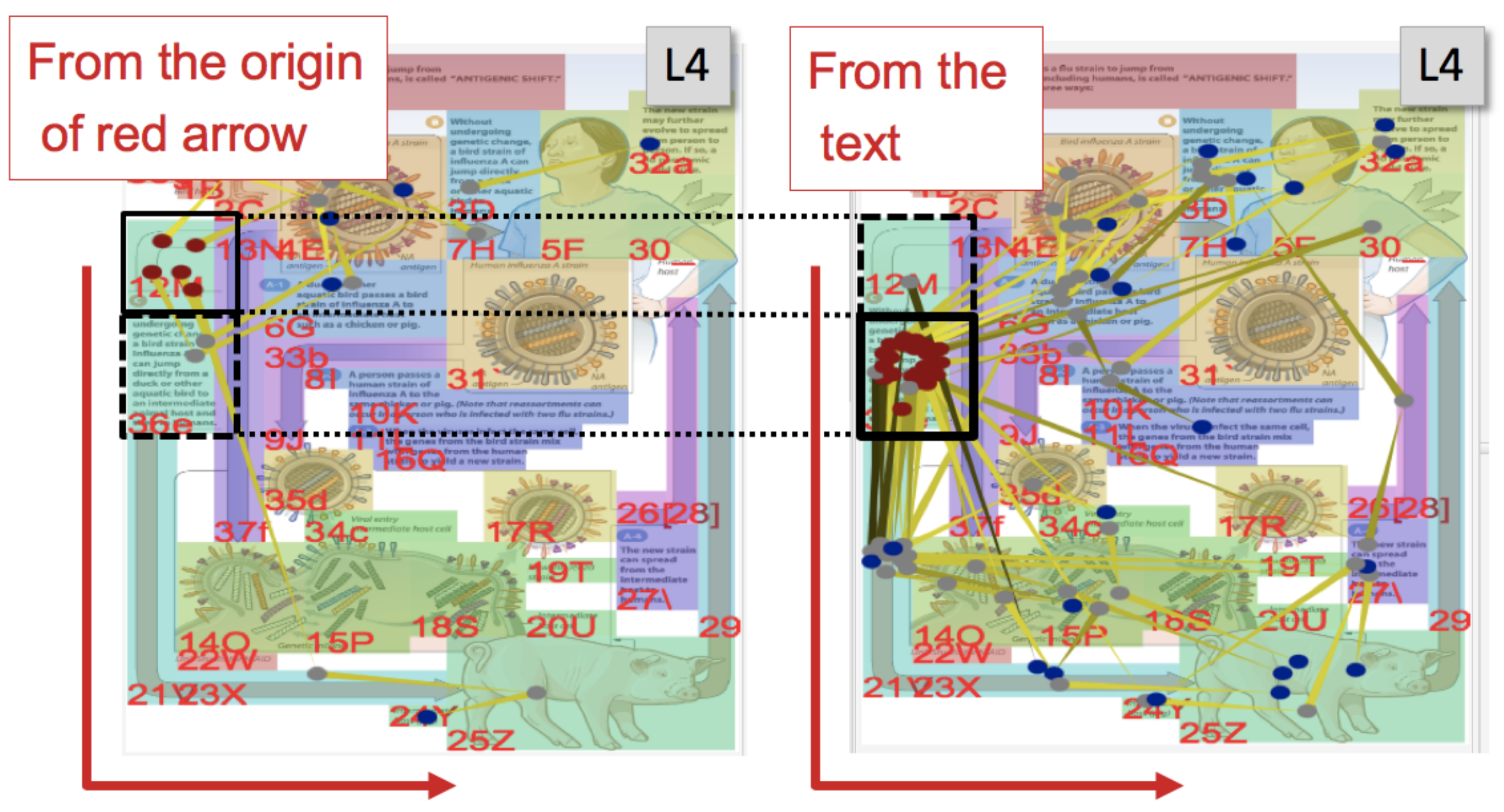}
\caption{
 Comparison of patterns extracted by 4-gram around the origin of the left red arrow.
 }
 \label{fig_antigenic1}
\end{figure}

Another arrow in the stimulus causes a different result.
Figure \ref{fig_antigenic1} shows patterns around the left red arrow.
In the left case, we extracted patterns that started from the origin of the red arrow.
As a result, there were no movements following the arrow completely.
Most of the patterns moved to the right blue route, even if they reached the following text on the red route.
One movement went to the lower side but passed the large picture without completely tracing the red arrow.
In the right figure, we visualized patterns that started from the text on the red arrow.
Many patterns followed the arrow at first; however, most of them turned to the picture next to the arrow on the halfway.
We suppose it is because the arrow is too long and turned on the halfway and therefore difficult to keep the attention of participants.

\begin{figure}[!h]
 \centering 
  \includegraphics[width=\columnwidth]{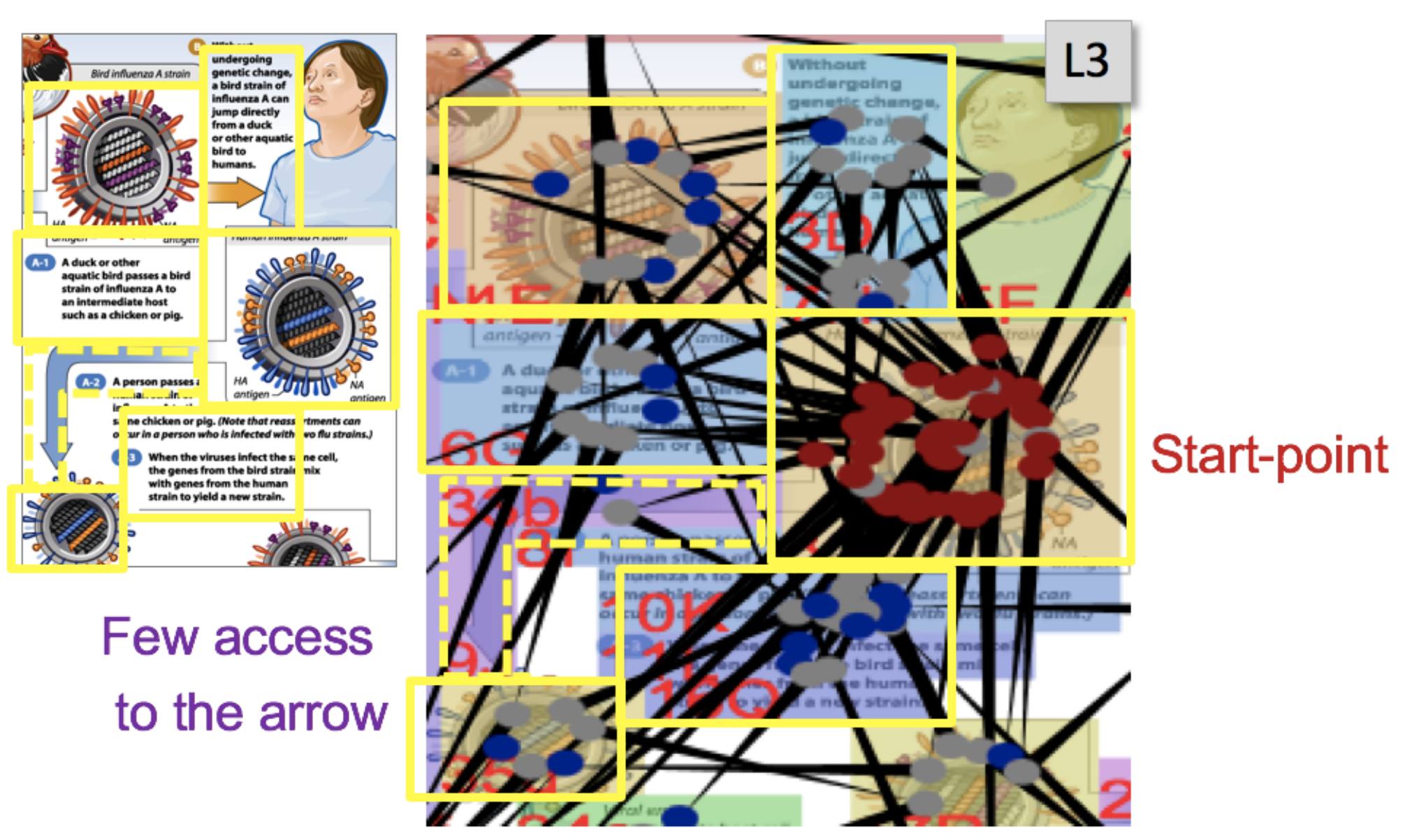}
 \caption{Patterns from the picture surrounded by some texts.}
 \label{fig_antigenic2}
\end{figure}

Figure \ref{fig_antigenic2} shows the visualization result of the center area.
We focused on one of the pictures and extracted patterns which started from the picture.
Texts and pictures around the focused picture collected many nodes, which denote frequent access to the texts and pictures.
However, there are few nodes at the arrow connected to the focused picture.
This result indicates that the participant tends to reach detailed information continuously even if they go out of the route indicated by an arrow.


\section{Discussion}
\label{sec_discussion}

In this section, we first summarize the experimental results and then discusses how we solved the two problems defined in Section 1.
Through the experiment, we verified the benefits of combining the two visualizations, drawing of scan-paths and displaying statistics of the pattern extraction.
Furthermore, we can find different characteristics by switching whether we firstly use one of the two visualizations.
``Bar chart view'' was helpful to find frequent and common transition patterns.
We used AOIs to describe the patterns so that we can understand what the participants focused on, not only where they looked at.
We did not have to memorize the specific positions of each AOI as we can visualize the selected patterns at the same time.
Using hierarchical AOIs and N-grams diversified the extracted patterns.
First, the hierarchical AOIs helped find both brief and fine behavior, particularly in Case A and B.
Another important feature is that users can construct the hierarchy of AOIs based on their relationships that users supposed.
Our technique sometimes visualized unexpected connections among AOIs that do not have similar contents indicated by yellow edges of the graphs because users did not suppose such AOIs had relationships.  
N-grams led to the variation of extracting patterns and showed different types of movements.
We could easily switch and compare patterns for visualization from short movements to long transitions in each use case.
%
Interactions on ``Bar chart view'' are useful to find which AOI is worth to focus on.
Meanwhile, click operations of AOIs on the stimulus realize the detail-on-demand interaction for already noticed AOIs.
It is helpful to understand usual orders of transitions by displaying the patterns that passed the selected AOIs.
Also, when we select patterns, the bars corresponding to the patterns got highlighted, and we could compare their frequencies. 
These mechanisms are especially effective with Case B: we could find interesting behaviors both from AOIs and the bar chart, as shown in Figures \ref{fig_uad0} and \ref{fig_uad1}. 
Like this, the combination of the two visualizations led to observe various features efficiently.

Then, we look back the two problems in Section 1.
The first problem is the visualization of patterns that pass multiple AOIs.
The suggestion of patterns to visualize by ``Bar chart view'' led to avoid visualizing too many movements.
We could visualize the long representative patterns that connect apart AOIs and find the participants tend to firstly focus on the upper area and then move to the lower area regardless of the contents on the stimulus.
The force-directed algorithm was also helpful to avoid overlapping of nodes and edges.
Notably, in the wide AOI, the nodes and edges scattered on the whole but partly gathered to connected directions.
Furthermore, our layout was effective to show a repetition of transitions between two AOI that often occur.
We visualized whole patterns even if they contain such repetition; however, the nodes got far off and avoided overlapping.
The second problem is on the comparison of multiple scan-paths.
``Bar chart view'' and ``Matrix view'' often helped find which participants had common behavior.
We could only visualize common or different patterns by selecting characteristic parts shown in ``Bar chart view''.
The concept of hierarchical AOIs was also effective to suppose how much the participants differ.
We found similar or non-similar participants in eight participants by ``Matrix view,'' and visualized characteristic transitions selected from ``Bar chart view'' in case A and B.

There are still some open issues.
On the force-directed graph layout, complicated results may occur in some severe cases, such as using very narrow AOIs or selecting many patterns to visualize.
The following two solutions are our ideas to solve this issue.
One is the algorithmic improvement for the layout; for example, adding repulsive forces to edges and an arrangement of swapping adjacent nodes.
The other is the improvement of the suggestion of patterns to visualize, such as more aggressive summarization of very similar patterns.
Related to the second solution, we also want to apply the summarization of the results of N-grams.
In our experiments, very few patterns are completely common when we selected more than 4-grams.
We should summarize some similar patterns, such as ``Including same AOIs but different order,'' `` Only single AOI is different'' if necessary.

\section{Conclusion}
\label{sec_conclusion} 

In this paper, we proposed a technique to extract common patterns in multiple eye tracking scan-paths and visualize them.
The comparison of N-grams and scan-paths are effective in finding characteristic patterns and understanding specific movements efficiently.
As future work, we would like to improve the layout of the graph and suggestion of patterns to visualize.


\acknowledgments{
This work was supported by Grant-in-Aid for JSPS Research Fellow Grant Number JP17J02298.
}

\bibliographystyle{unsrt}


\end{document}